\documentclass[aps,twocolumn,prd,preprint,groupedaddress,10pt]{revtex4}

\usepackage{graphicx}
\usepackage{amssymb}
\usepackage{amsmath}

\begin{document}


\title{High speed collision and reconnection of Abelian Higgs strings in the deep type-II regime}


\author{G.J. Verbiest}
\email[]{Verbiest@physics.leidenuniv.nl}
\affiliation{Instituut-Lorentz for Theoretical Physics, Leiden, The Netherlands}
\author{A. Ach\'{u}carro}
\email[]{Achucar@lorentz.leidenuniv.nl}
\affiliation{Instituut-Lorentz for Theoretical Physics, Leiden, The
  Netherlands}
\affiliation{Departamento de F{\'\i}sica Te\'orica, Universidad del
  Pa{\'\i}s Vasco UPV-EHU, Bilbao, Spain}


\date{\today}

\begin{abstract}
  We study the high speed collision and reconnection of
  Abrikosov--Nielsen--Olesen cosmic strings in the type-II regime of
  the Abelian Higgs model, that is, scalar-to-gauge mass ratios larger
  than one. Qualitatively new phenomena such as multiple reconnections
  and clustering of small scale structure have been observed in the
  deep type-II regime and reported in a previous paper, as well as the
  fact that the previously observed ``loop'' that mediates the second
  intercommutation is only a loop for sufficiently large mass ratios.
  Here we give a more detailed account of our study, which involves 3D
  numerical simulations with the parameter $\beta =
  m_{scalar}^2/m_{gauge}^2$ in the range $1 \leq \beta \leq 64$, the
  largest value simulated to date, as well as 2D simulations of
  vortex-antivortex head-on collisions to understand their possible
  relation to the new 3D phenomena.

  Our simulations give further support to the idea that Abelian Higgs
  strings never pass through each other, even at ultrarelativistic
  speeds, unless this is the result of a double reconnection; and that
  the critical velocity for double reconnection goes down with
  increasing mass ratio, but energy conservation suggests a
  lower bound around 0.77 c. We discuss the qualitative change in the
  intermediate state observed for large mass ratios. We relate it to a
  similar 
  change in the outcome of 2D vortex--antivortex collisions in the
  form of radiating bound states, whereas  we find no evidence
  of the back-to-back reemergence reported in previous studies.

  In the deep type-II regime the angular dependence of the critical
  speed for double reconnection does not seem to conform to the
  semi-analytic predictions based on the Nambu-Goto approximation.  We
  can model the high angle collisions reasonably well by incorporating
  the effect of core interactions, and the torque they produce on the
  approaching strings, into the Nambu--Goto description of the
  collision. An interesting, counterintuitive aspect is that the
  effective collision angle is smaller (not larger) as a result of the
  torque. Our results suggest differences in network evolution and
  radiation output with respect to the predictions based on
  Nambu--Goto or $\beta = 1$ Abelian Higgs dynamics.
\end{abstract}


\maketitle









\section*{Introduction}

The discovery of cosmic strings, first proposed by Kibble
\cite{Kibble} would revolutionize our understanding of particle
physics at the extremely high energies present in the very early
Universe. It could signal a `superconducting' phase transition and
give information on the particle interactions before the transition,
or it might provide the first evidence of superstring theory.  So far
there is no evidence of their existence, but
strings can lead to a wealth of detectable astrophysical phenomena and
there is an increasing number of surveys and searches looking for
observable signatures. These include Cosmic Microwave Background (CMB)
anisotropies, gravitational lensing, wakes, gravitational radiation
(GW), cosmic rays and gamma ray bursts, among others. Gravitational
effects are determined by the adimensional parameter $G \mu$ which for
most models of cosmological interest falls in the range $10^{-13}-
10^{-6}$ (we use units with $\hbar = c = 1$ throughout; $G$ is
Newton's constant and $\mu$ the mass per unit length of the
strings). Strings with higher mass per unit length are already ruled
out by these observations, see the classic reviews
\cite{VilenkinShellard,
  HindmarshKibble,Polchinski:2004ia,Davis:2005dd} and the more recent
updates \cite{CopelandKibble,Sakellariadou:2009ev,Achucarro:2008fn,
  Ringeval:2010ca, Copeland:2011dx, Hindmarsh:2011qj}, and references
therein.

The formation of cosmic strings and superstrings is a generic outcome
of cosmological phase transitions\cite{Kibble} and of some
inflationary models, in particular those based on Grand Unified
theories\cite{Jeannerot,JeannerotRocherSakellariadou}. More recently
it has been appreciated that some models of brane inflation could also
lead to cosmic superstrings \cite{MajumdarDavis,SarangiTye}.  Once
formed, a string network is expected to reach a scaling solution in
which statistical properties such as the distance between strings or
the persistence length become a fixed fraction of the horizon size
(the age of the Universe). The energy density in strings decreases
with time but the expansion of the Universe pumps energy into the
string network --by increasing the contribution of long strings-- and
this is balanced by energy losses to radiation. If these are efficient
enough, the contribution from the strings to the energy density of the
Universe remains a small, constant fraction of the dominant form of
energy (matter or radiation) and is potentially observable. Radiation
is emitted by oscillating loops, formed when a string self-intersects,
and in bursts. The latter are produced by cusps (sections of the
string which acquire near-luminal speeds), by the final stages of
collapsing loops and, to a lesser extent, by kinks created when
strings reconnect.  The reconnection, or {\it intercommutation} of
strings is therefore an essential process that determines and
maintains the long-term scaling
behaviour of the string network.\\

In this paper, a companion to \cite{us}, we focus on the high speed
intercommutation of Abelian Higgs strings in the deep type-II
regime. The terminology is borrowed from superconductors, where type-I
(type-II) indicates a critical parameter $\beta<1$ ($\beta>1$);
$\beta$ is the ratio of scalar to gauge excitation masses, squared
(see below). An important difference between type-I and type-II is the
interaction energy between parallel vortices, which is attractive for
type-I and repulsive for type-II\cite{JacobsRebbi}. Cosmic string
intercommutation was first investigated numerically by Shellard for
global strings \cite{Shellard} and later by Matzner for type-II $\beta
= 2$ Abelian Higgs strings \cite{Matzner}. They pointed out that in
ultra-relativistic collisions there is a critical center-of-mass
velocity $v_c$ (depending on the collision angle) beyond which strings
pass through: a loop forms, expanding rapidly from the collision
point, that catches up with the reconnected strings and produces a
second intercommutation. A more recent study \cite{Putter} focused on
double reconnection and showed that it proceeds differently in type-I
and type-II strings, with the loop only forming in type-II
collisions. In fact, one of the main results in \cite{us} that we
elaborate on here is that the loop only forms in deep type-II ($\beta
>> 1$) collisions. Although we do not determine the precise value of
$\beta$ for which the transition occurs, it is somewhere between
$\beta = 8$ and $\beta = 16$. For $\beta \geq 16$ we see the loop
forming, while type-II collisions with $\beta \leq 8$ produce instead
a blob of radiation that can look like a loop but is not (see later
sections and \cite{us}). Another important point is the angular
dependence of $v_c$ as a function of the collision angle. In
ref. \cite{Putter} it was shown that it is dictated by the geometry
and speeds of the strings after the first reconnection, which can be
calculated in the Nambu-Goto approximation.  In the deep type-II
regime we expect core interactions to play an increasingly important
role,
even before the collision, and this is indeed what we will report here.\\


The simulations in \cite{Putter} had decreasing resolution with
increasing $\beta$, and only explored two values of $\beta$ in the
type-II regime ($\beta = 8,32$), but the results suggested that the
critical velocity for the second reconnection would go down as a
function of $\beta$.  This is interesting because one of the
distinguishing features of cosmic superstrings, as opposed to the
Abrikosov--Nielsen--Olesen (ANO) strings \cite{Abrikosov,
  Nielsen:1973cs} considered here, is their low intercommutation
probability $P \sim 10^{-3} - 10^{-1}$ \cite{Jackson:2004zg}, which
leads to different scaling properties, in particular to denser
networks (although in an expanding background the effect is weaker
than the $\rho t^2 \sim 1/P$ dependence one might expect for the
density $\rho$ at cosmic time $t$
\cite{AvgoustidisShellard}). However, if the critical velocity of
strongly type-II Abelian Higgs strings decreases to the extent that it
becomes comparable to the average velocity of the network, the {\it
  effective} intercommutation probability could be
much less than one. This was one of the motivations behind the present study.\\




Our results confirm the claim of \cite{Putter} that the critical
velocity for the second reconnection goes down as a function of
$\beta$, although energy conservation suggests this decrease
cannot go on indefinitely. We will return to this point later.
Furthermore, while studying the critical velocity, we found multiple
intercommutations. That is, processes where the strings exchange ends
three times or more. We only found multiple reconnections for $\beta
\geq 16$. We will show that this is related to a
qualitative change in the nature of the intermediate state, from
localized radiation to a string loop,
which determines the process leading to the second intercommutation. For $1 < \beta \leq 8$ we found that the previously reported loop is just an expanding blob of radiation, while for $\beta \geq 16$ we find a topological loop rapidly expanding from the collision point.\\

An interesting question that was not addressed in \cite{us} is whether
these multiple reconnections and the blob-to-loop transition could be
related to, for instance, two-dimensional bound states of the
vortex-antivortex system or whether they are a purely
three-dimensional effect. In order to shed light on this connection we
also simulated the high speed head-on collision of a vortex and an
antivortex in 2D.  Abelian Higgs vortex-antivortex scattering in two
dimensions was studied years ago by Myers, Rebbi and
Strilka\cite{MyersRebbiStrilka} (see also
\cite{MoriartyMyersRebbi}). They reported that, beyond a certain
critical speed, the vortices reemerge and the way in which they
reemerge depends on $\beta$: for $\beta \leq 4$ they bounce back and
for $\beta \geq 8$ they pass through (and for velocities below the
critical speed the pair annihilates into radiation). For large $\beta$
we agree with their results: we found that, for $\beta \geq 6.4$, the
vortices reemerge as if they have passed through. However, for $\beta
\leq 6.2$ we find no evidence of backscatter; instead, we always find
a bound radiating state. The energy profile looks as if the
vortex-antivortex have passed through but a closer look at the
magnetic field shows they are just fluctuations.  They cannot escape
each other's influence and radiation keeps being emitted from the area
around the collision point until all the energy has been radiated
away.  As we will discuss, these two behaviours are consistent with
the change we observe in the intermediate state in 3D collisions. The
``pass through'' behaviour at high $\beta$ corresponds to a 3D loop
while the 2D radiating bound states at low $\beta$ correspond to the
3D ``blob''. On the other hand, the relation with multiple
reconnections is more subtle, and it appears that the number of
reconnections is mainly a three-dimensional effect.

\section*{A few comments on cosmic string intercommutation}

String intercommutation poses some interesting puzzles from a
theoretical point of view.  ANO strings are topological, they cannot
break. When two strings segments collide there are, in principle, three
possible outcomes:\\

a) They can exchange partners and reconnect (intercommute). The
reconnected strings have a slightly shorter length. This is the
default outcome and, as we shall discuss, seems to occur {\it always}
when Abelian Higgs strings meet.\\

b) In near parallel collisions of Type-I strings at low velocity, the
attractive interaction makes them stick together, e.g. type-I Abelian
Higgs strings\cite{BettencourtKibble,BettencourtLagunaMatzner}. The
network then has junctions between strings with different winding
numbers (and different energy per unit length). String networks with
junctions have been extensively studied analytically in the Nambu-Goto
approximation\cite{Copeland:2006if,CopelandKibbleSteer} and in this regime they
give an extremely good fit to numerical simulations \cite{Salmi}.\\


c) The strings can simply pass through each other.  The relative
probability of outcomes a) and c) is an important distinguishing
feature of ANO strings versus cosmic (super)strings.  The
intercommutation of fundamental superstrings, and of D-strings, is a
quantum process and, as such, has some probability of not
happening\cite{Polchinski:1988cn}. This is also an expected feature of
higher dimensional models --in which the apparent collision of the
strings can be a four dimensional illusion, the strings are actually
not intersecting in the higher dimensional space-- or models with
extra internal degrees of freedom
\cite{Polchinski:2004ia,Hashimoto:2005hi}. Cosmic superstrings have been
shown to give reconnection probabilities $(P)$ as low as $P \sim
10^{-3} - 10^{-1}$\cite{Jackson:2004zg}.\\

An interesting observation is that one would naively expect outcome c)
to be the result of any sufficiently fast collision of ANO strings,
since the natural timescales of the microscopic fields can be much
slower than the string crossing time\cite{CopelandTurok}. In the case
of domain walls, for instance, the scalar field profiles pass through
each other undistorted in ultrarelativistic collisions
\cite{Giblin:2010bd}. The head-on collision of a vortex and an
antivortex in 2D at ultrarelativistic speeds is also known to result
in the two passing through each other
\cite{MyersRebbiStrilka}. However this ``free passage'' is not
observed in 3D string collisions.
In the Abelian Higgs model, all (numerical) evidence to date
\cite{Matzner, Putter,us} points to the conclusion that $P=1$:
Abrikosov-Nielsen-Olesen (ANO) strings with unit winding {\bf always
  reconnect at least once, even at ultrarelativistic speeds} (even in
case b) there is some evidence that the strings will first reconnect
and then settle into a junction \cite{Salmi}).  What is observed,
instead, is that beyond a certain critical collision velocity $v_c$,
the strings may reconnect a second time and effectively pass through
with some distorsion. In terms of network evolution this is as if the
strings have not reconnected, and one can talk of an {\it effective}
intercommutation probability $P_{eff}$ being less than 1.  An
important question is to model the dependence of the critical velocity
for non-reconnection as a function of collision speed and angle, and
to understand its dependence with $\beta$.
\\

\begin{figure}
\includegraphics[width=85mm]{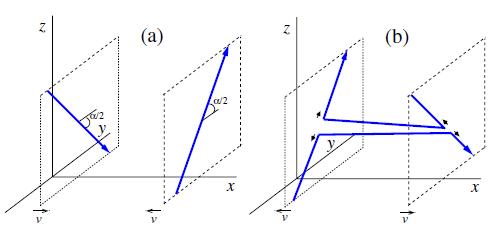}
\caption{From ref. \cite{Putter} (a) Initial positions and orientations of
  the strings in the center of mass frame. The strings lie in the $x =
  const.$ planes and approach each other with speed $v$. The arrows
  indicate the orientations of the strings, which form an angle
  $\alpha$. (b) The configurations after one intercommutation. If $v
  \sim c$, the kinks' motion along the strings is negligible and the
  connecting horizontal segments are practically antiparallel and
  immobile, making a second interaction possible.\label{start_config}}
\end{figure}

At lower speeds, an interesting class of analytic results on intercommutation
is based on the moduli space approximation \cite{Manton:1981mp}. In
the Bogomolnyi limit ($\beta = 1$) \cite{Bogomolnyi}, the moduli space
or geodesic approximation gives a very good description of slow
vortex--vortex collisions in two dimensions and predicts the observed
right-angle scattering\cite{Ruback}. Although these arguments are only
valid at low speeds and in the absence of core interactions, they lead
to the expectation that in three dimensional collisions reconnection
is inevitable. Consider the configuration in
fig. \ref{start_config}a), two strings colliding with angle $\alpha$
and velocity $v$. If the incoming strings lie in the $yz$ plane and
move in the $x$-direction, the collision looks in the $xz$ plane like
two vortices that will scatter at $90^0$
\cite{Samols,MoriartyMyersRebbi,Ruback}
while in the $xy$ plane it is a vortex-antivortex collision, which we expect will annihilate. In 3D the energy from this annihilation goes into creating the connecting segments and the rest is emitted as radiation. \\

This idea has been extended to other models such as D-strings
\cite{HananyHashimoto} and non-Abelian strings
\cite{Hashimoto:2005hi,Eto:2006db}). In many cases intercommutation is
expected with probability one in the regime of validity of the moduli
space approximation, which --as we just emphasized-- requires two
conditions: low speed, so higher order frequency modes are not
excited, and negligible core interactions (such as in near parallel
collisions not far from
the Bogomolnyi limit). \\

It is worth stressing that even if these conditions are satisfied
before the collision, they will not hold after intercommutation
because the ``new'' portions of string that are generated between the
receding strings are necessarily almost antiparallel around the point
of collision, and their core interactions are crucial in understanding
what happens next. This is probably the reason why a prediction for
the angular dependence of the critical speed in \cite{HananyHashimoto}
fails to agree with the data from numerical simulation. This
prediction is based on an energy argument which does not take into
account the interaction between the string cores, and this
approximation fails immediately {\it after} the collision.
\\



By using the thin string approximation and taking into account the
effect of core interactions after the collision, ref. \cite{Putter}
obtained a semi-analytic expression for the critical velocity for
double reconnection $v_c$ as a function of the collision velocity $v$
and angle $\alpha$. This works quite well for type-I and moderately
type-II ANO strings.  Consider the situation in figure
\ref{start_config}b) after the first intercommutation. Within the
Nambu-Goto approximation --thus, ignoring core interactions-- we can
express the angle $\delta$ between the horizontal segments and the
speed $w$ at which the horizontal segments move apart in terms of $v$
and $\alpha$:

\begin{equation}
\label{w_crit}
w = \sin (\alpha/2)/\gamma(v)
\end{equation}
\begin{equation}
\label{delta_crit}
\cos(\delta/2) = \frac{\cos (\alpha/2)/(v\gamma(v))}{\sqrt{1+ (\cos (\alpha/2)/(v\gamma(v)))^2}}
\end{equation}

Notice that for high collision speeds the horizontal segments are
almost antiparallel, $\delta \sim \pi$ and they move slowly $w \sim
0$.  If they are not receding very fast, the anti-aligned segments
will be attracted to each other and will annihilate, causing the
strings to reconnect a second time. In the unphysical limit of a
collision at the speed of light, the bridging segments would not move
at all and would be antiparallel, so the second reconnection is
expected with probability one. The angular dependence of the threshold
speed $v_c$ is then found with equations (\ref{w_crit}) and
(\ref{delta_crit}) if one assumes
that reconnection will happen below a threshold ``escape'' speed
$w_t$
and above  a threshold angle $\delta_t$, close to antiparallel. The threshold values $w_t$ and $\delta_t$ are left as free parameters and determined by the best fit to the numerical data. While this model works well near the Bogomolnyi limit $\beta \sim 1$, we will show that it does not work so well in the deep type-II regime studied here, indicating that core interactions before the collision are also important. We will return to this point in the discussion.  \\

Equation (\ref{w_crit}) shows that fast collisions will produce very
slowly moving connecting segments. Conversely, a slow collision
usually creates a highly curved region after intercommutation which
will accelerate under its own tension and acquire large speeds. This
is one of the mechanisms that helps the string network maintain a
typical speed $<v^2> \ \sim 0.5$,  even in an
expanding Universe.\\

\section*{Simulations}

The Abelian Higgs model is the relativistic version of the
Ginzburg-Landau model of superconductivity.
It is described by the lagrangian
\begin{equation}
\label{NM01}
\mathcal{L} = (\partial_\mu + i e A_\mu)\phi(\partial^\mu - i e A^\mu)\phi^\dagger - \frac{1}{4} F^{\mu\nu} F_{\mu\nu} - \frac{\lambda}{4}(|\phi|^2 - \eta^2)^2
\end{equation}
where $\phi$ is a complex scalar field and $A_\mu$ is a U(1) gauge
field with field strength $F_{\mu\nu} = \partial_\mu A_\nu
- \partial_\nu A_\mu$ ($\mu, \nu = 0,1,2,3$).\\

The ground state has $|\phi| = \eta$ and zero electric and magnetic
field. The fluctuations about this vacuum define two mass scales: the
scalar excitations have $m_{scalar} = \sqrt \lambda \eta$ and the
gauge field excitations, $m_{gauge}=\sqrt 2 e \eta$. Classically, the
only relevant parameter in the dynamics is their ratio, $\beta =
(m_{scalar}/m_{gauge})^2 = \lambda / 2e^2$, which also characterizes
the internal structure of the
ANO vortices.\\

Magnetic cores repel and scalar cores attract, so the interaction
between vortices is determined by which of these cores is the largest:
parallel ANO vortices repel for $\beta > 1$ and attract for $\beta <
1$. The Bogomolnyi limit $\beta = 1$ is a critical value where both
effects cancel and parallel vortices do not interact. In this paper we
are interested in the $\beta > 1$ regime, analogous to a type-II
superconductor, and in this case the vortices have an inner ``scalar''
core of radius $\sim m_{scalar}^{-1}$ in which the scalar field
departs from its vacuum value and vanishes at the center. This is
surrounded by a larger, ``gauge'' core of radius $\sim m_{gauge}^{-1}$
where the magnetic field is non-zero.  The repulsive interaction
produces a torque that tends to anti-align two colliding
strings. This will play an important role later.\\



Here, as in \cite{us}, we follow the numerical strategy of \cite{Matzner,Putter}: we use a lattice
discretization and place a superposition of two oppositely moving ANO
strings on a three dimensional lattice. This configuration is evolved
using a leapfrog algorithm. The initial configuration is determined by
two parameters: the center-of-mass speed $v$ of the strings when they
are far apart and the angle $\alpha$ between them (every collision can
be brought to this form by an appropriate Lorentz transformation \cite{Shellard}).\\

We also impose  ``freely moving'' boundary conditions:
after each round the fields inside the box are updated using the
equations of motion, and the fields on the boundaries are calculated
assuming the strings move unperturbed and at
constant speeds at the boundaries.\\

All 3D simulations were done on a $200 \times 200 \times 400$ grid.
Unless otherwise stated, we use a lattice spacing $a = 0.2$ and time
steps $\Delta t = 0.02$, so the Courant condition (here $\Delta t \leq
a/\sqrt{3}$) holds. Some simulations were repeated with a = 0.1, in particular those with $\beta = 4$, to confirm the results.\\

Our simulations are optimized for
the deep type-II regime.  By solving the two-dimensional, static
vortex equations one finds that in a static straight cosmic string
about half of the potential energy in the scalar core is contained within a
radius $\sqrt 2 m_{scalar}^{-1} f(\beta)$, where $f$ is a slowly
varying function with $f(1)= 1, f(64) = 1.4$. Lorentz-contraction
gives an extra factor $\gamma(v)^{-1}$ in the direction of approach,
with $\gamma(v) = 1 / \sqrt{1 - v^2}$.  This is the smallest length
scale that has to be resolved.
Without loss of generality we take $\lambda = 2, \ \eta = 1$, which
ties the unit of length (and time) to $m_{scalar}^{-1} = 1
/ \sqrt 2$.  The scalar core is resolved by at least three lattice
points up to a center of mass speed of $v \approx 0.94-0.96$, which is
indicated explicitly in the diagrams in figure \ref{vcrit_panels} (it
is higher in the $\beta = 4$ simulations because those have $a=0.1$).
The initial string separation is fixed to
$5\sqrt{2\beta} \gamma(v) m_{scalar}^{-1} = 5\sqrt{\beta}
\gamma(v)$. This would be about five times the actual core radii for
$\beta = 1$ but as $\beta$ increases, the core sizes increase and for
large $\beta$ one has to check that the gauge cores do not overlap
in the initial configuration.
For $\beta = 64$ the overlap in total energy from the tails of the
gauge cores is less than 1\% when $\gamma(v) = 1$, that is, when
calculated on static vortices.
\\
Finally, the two-dimensional simulations of vortex-antivortex head-on
collision and reemergence were done on a 800x800 grid with a lattice
spacing of 0.1 and $\Delta t = 0.02$. In this way we resolve the
vortex cores with at least 3 points up to a speed of $v
\approx 0.985$. The initial separation is the same as in the 3D
simulations. We used absorbing boundary conditions. In all simulations
(2D and 3D) energy is conserved to better than 5\% until the radiation
hits the boundary
(which determines the dynamical range).\\

\section*{Results}

We simulated the collision of cosmic strings at $\beta = 1, 3.9, 4.0,
4.1, 8, 16, 31, 32, 33, 49$ and $64$ for various speeds $v$ and angles
$\alpha$ to find the threshold velocity above which the strings
effectively pass through each other. The results for selected values
of $\beta$ are shown in figure \ref{vcrit_panels}.  Results for $\beta
= 3.9$ and $4.1$ were qualitatively similar to $\beta = 4$; also,
results for $\beta = 31, 32$ and $33$ were qualitatively similar.
Some salient features were already reported in reference
\cite{us}.\\

The first thing that is apparent in fig. \ref{vcrit_panels} is that
the minimum critical velocity for a second reconnection goes down as a
function of $\beta$, in agreement with what was observed in
\cite{Putter}. The dependence of this lowest critical velocity $v_{c,
  min}$ with $\beta$ and the range of collision angles for which it is
observed is seen in table \ref{v_min}.\\

\begin{table}
\begin{center}
\begin{tabular}{|c|c|c|}
$\beta$ & $v_{c, min}$ & $\alpha$ \\
\hline
4 & 0.92 & $42^0$ \\
8 & 0.92 & $62^0$ \\
16 & 0.90 & $90^0-110^0$ \\
32 & 0.88 & $80^0-120^0$ \\
49 & 0.88 & $85^0-125^0$ \\
64 & 0.86 & $105^0-120^0$
\end{tabular}
\end{center}
\caption{The lowest value of the critical velocity for double reconnection as a function of $\beta$, and the collision angles at which it is observed. The range in $\alpha$ indicates the possible existence of a plateau in the critical velocity.}
\label{v_min}
\end{table}



Secondly, there is a new phenomenon of multiple intercommutations, which appears to be related to a change in the nature of the intermediate state from a non-topological blob of radiation to a loop.\\

\begin{figure}
\begin{tabular}{cc}
\includegraphics[height = 40mm,width=40mm]{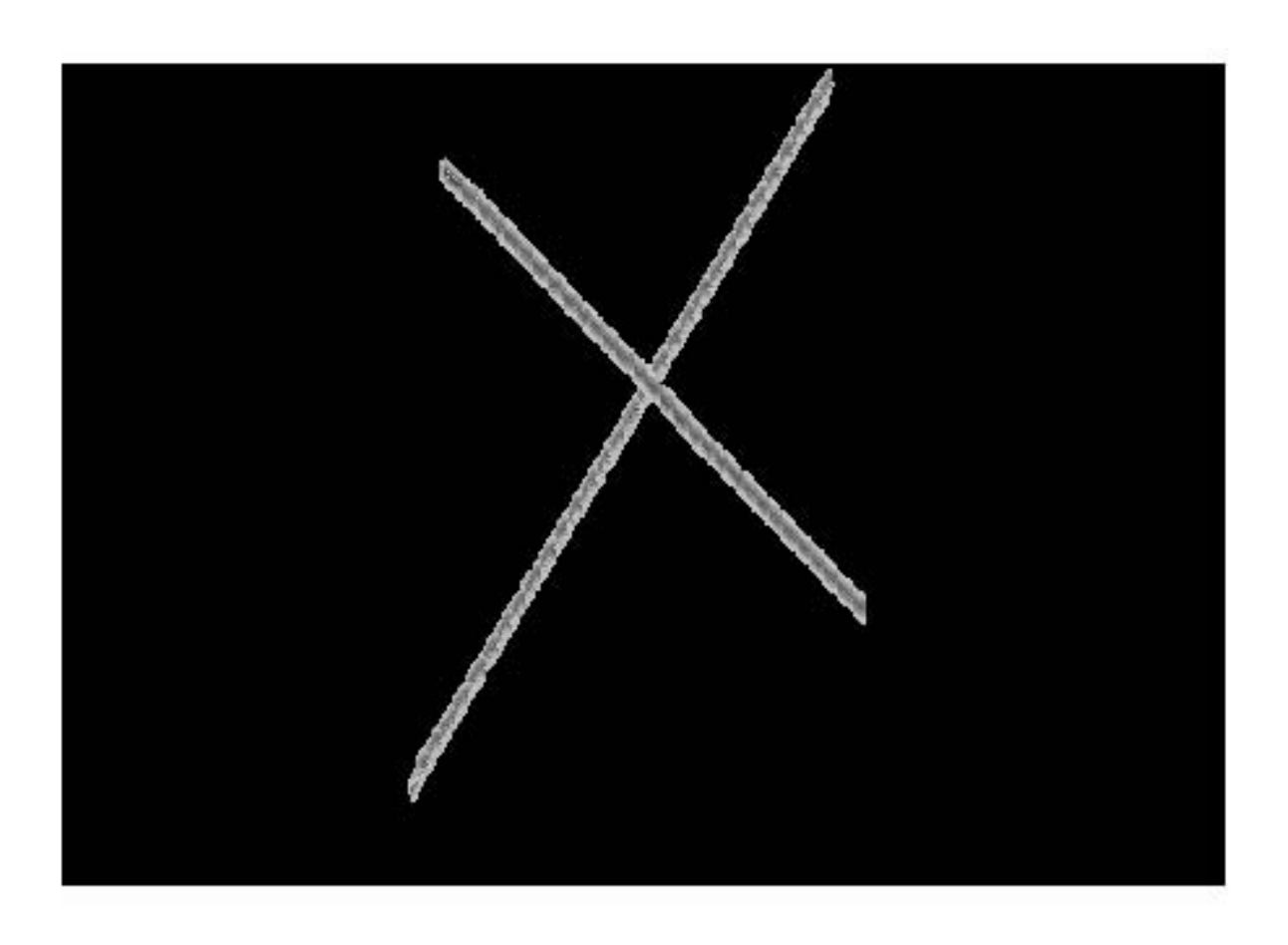} &
\includegraphics[height = 40mm,width=40mm]{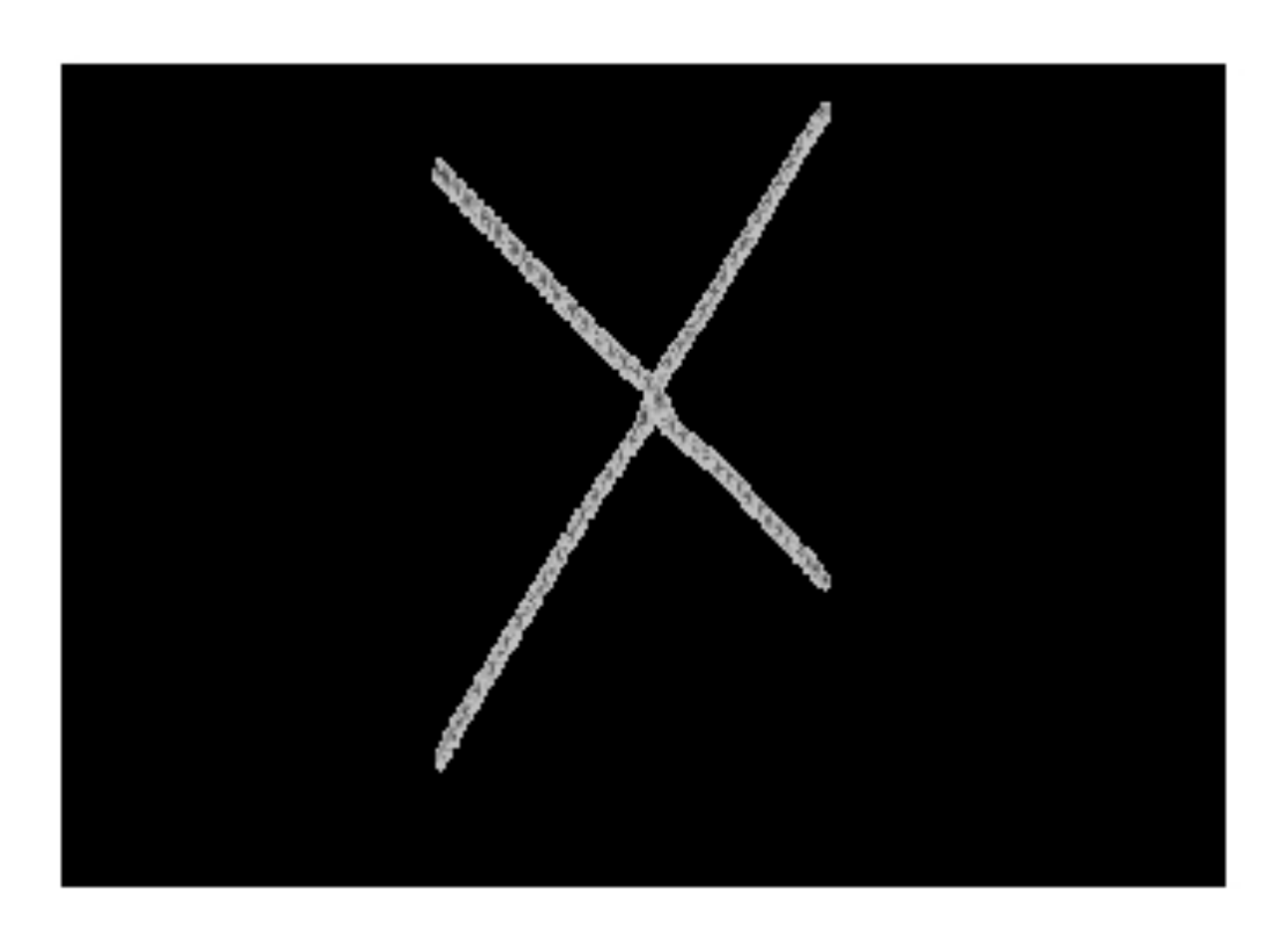} \\
\includegraphics[height = 40mm,width=40mm]{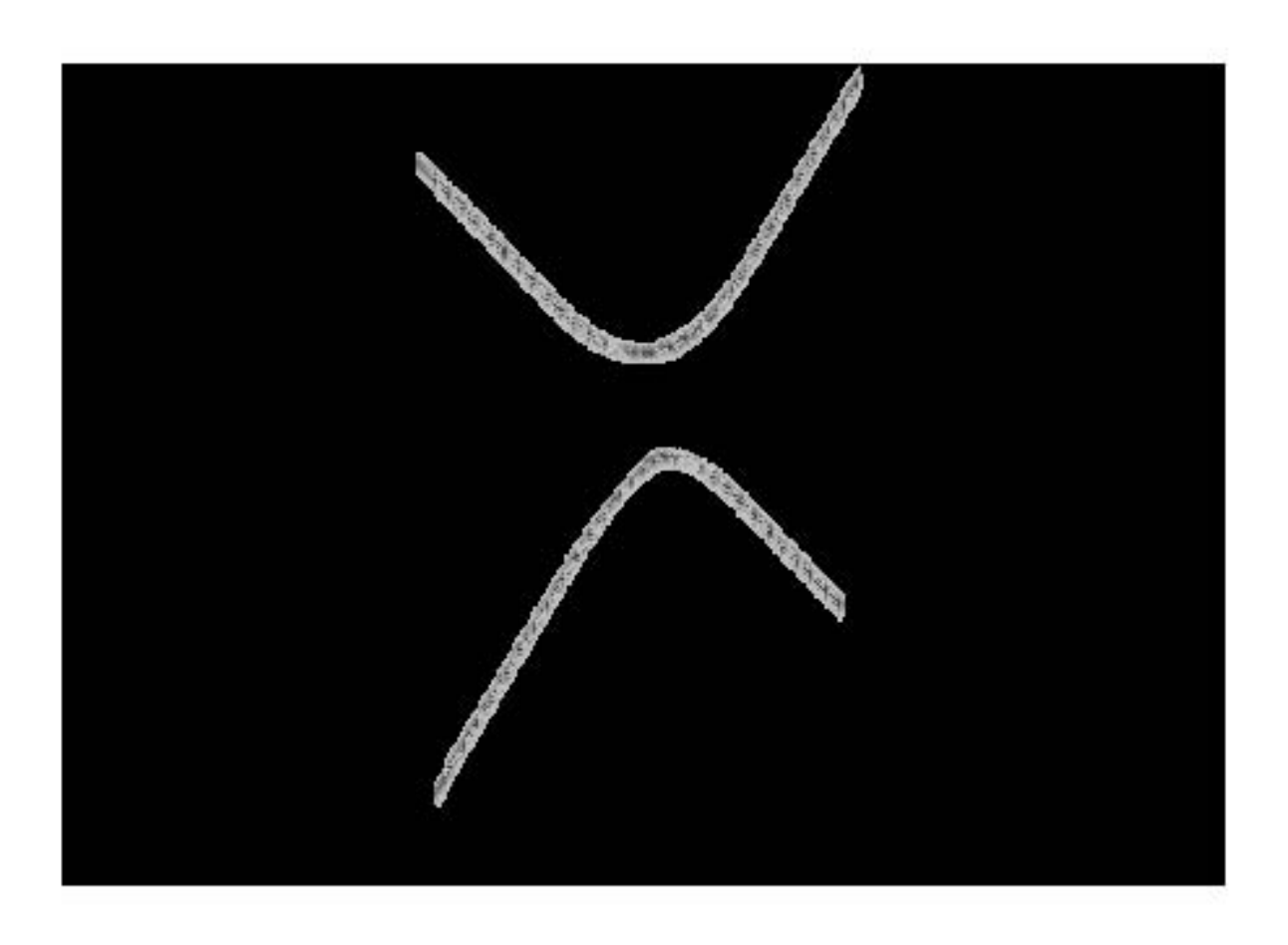} &
\includegraphics[height = 40mm,width=40mm]{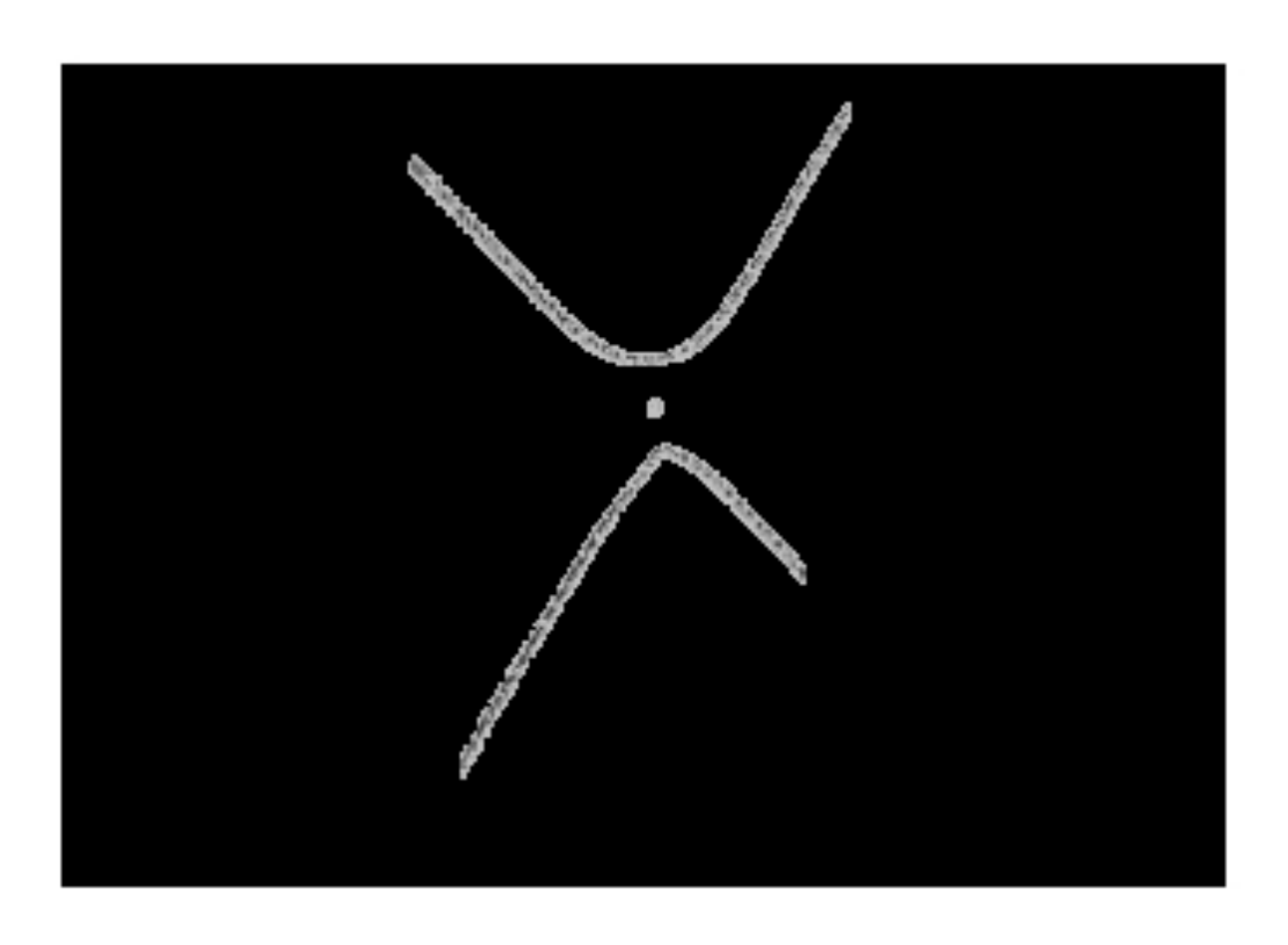} \\
\includegraphics[height = 40mm,width=40mm]{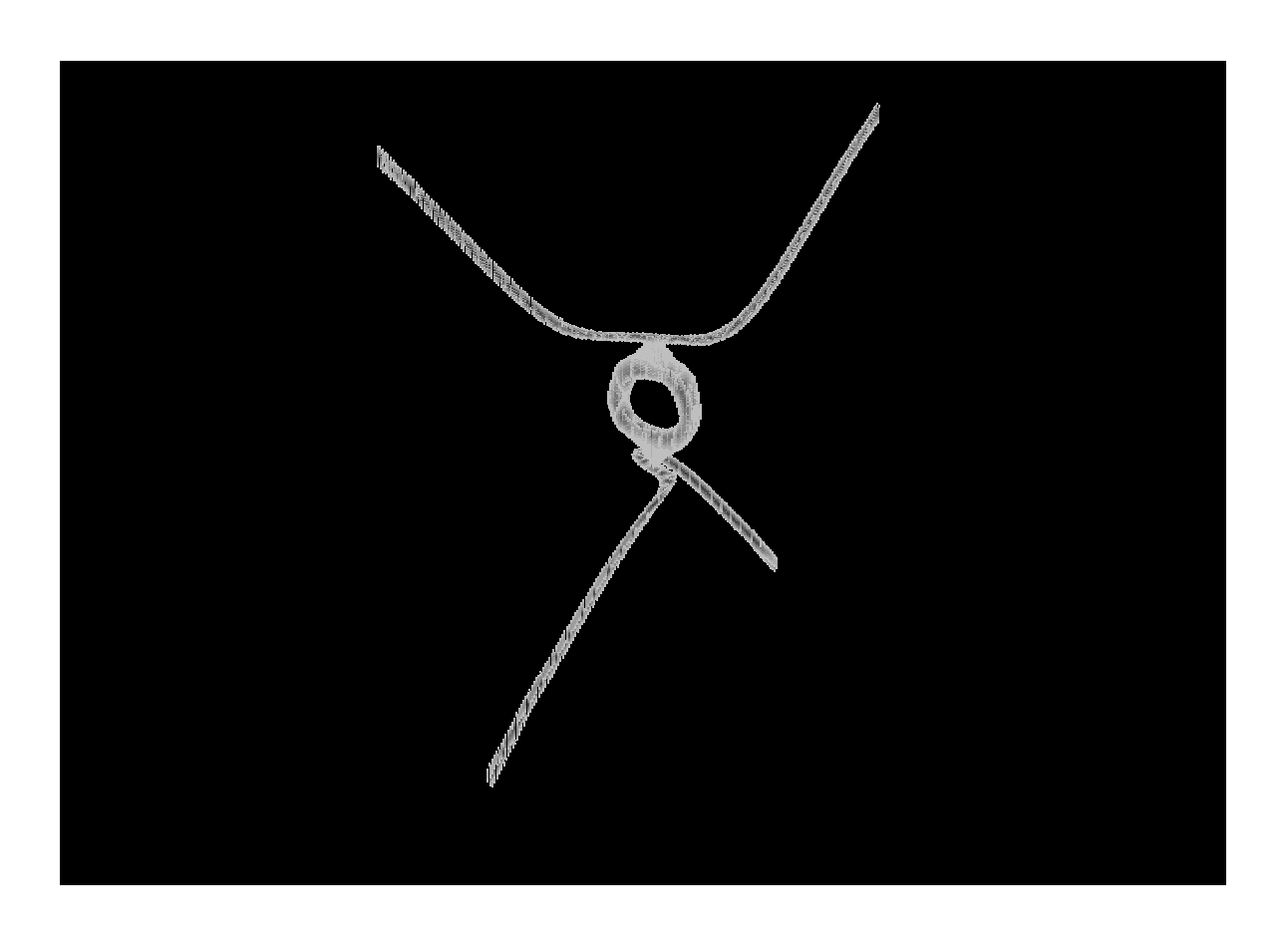} &
\includegraphics[height = 40mm,width=40mm]{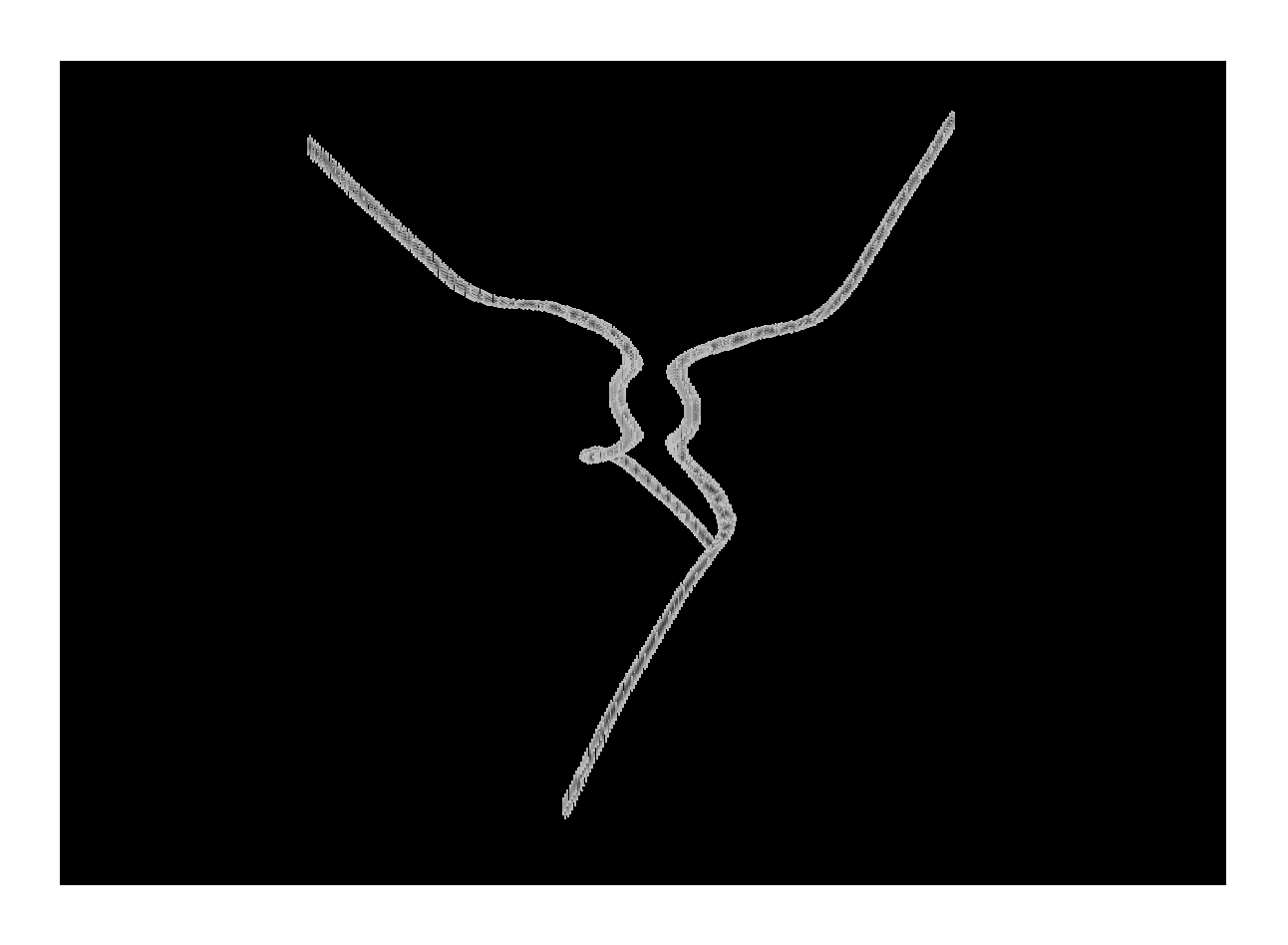} \\
\includegraphics[height = 40mm,width=40mm]{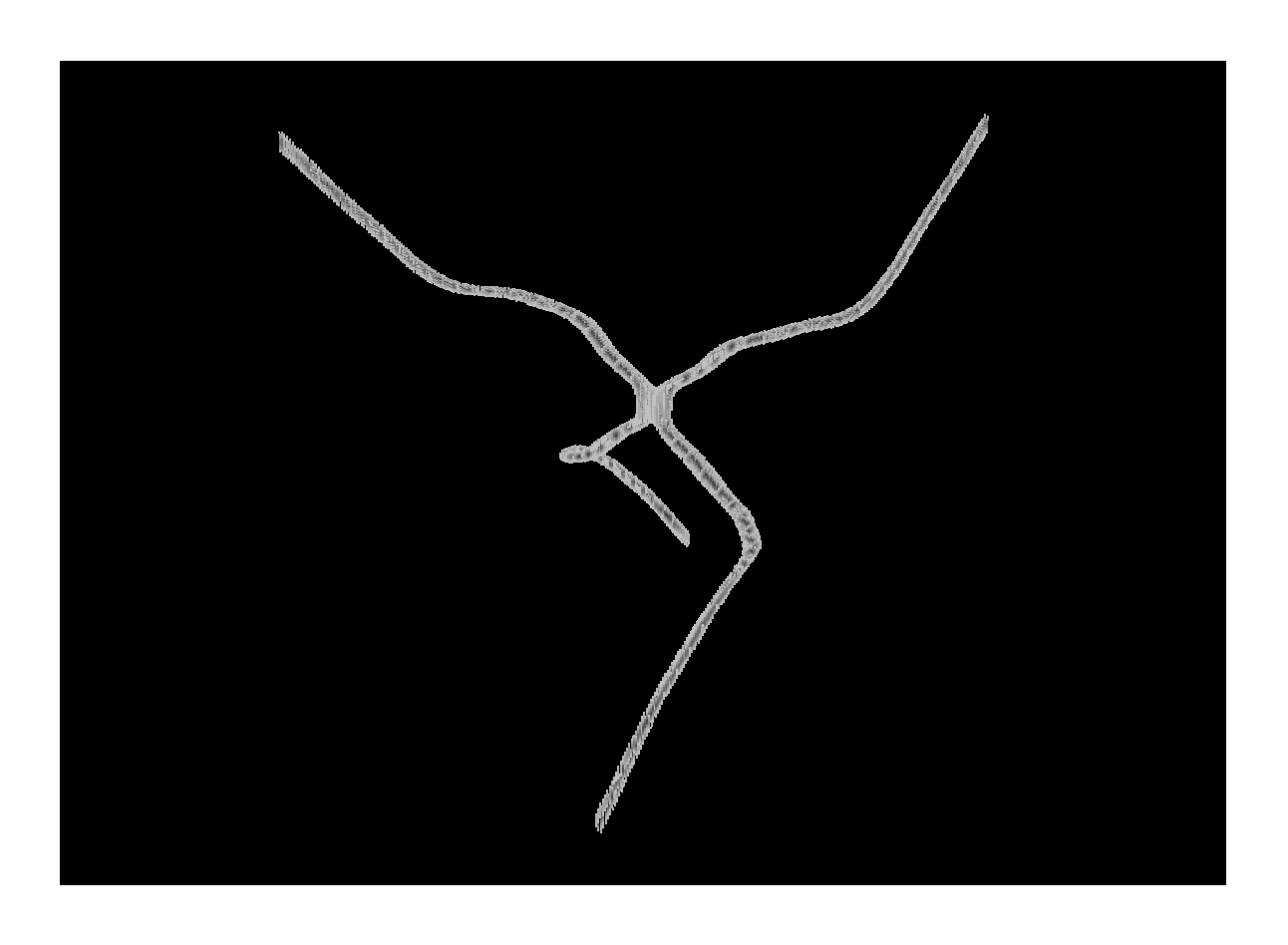} &
\includegraphics[height = 40mm,width=40mm]{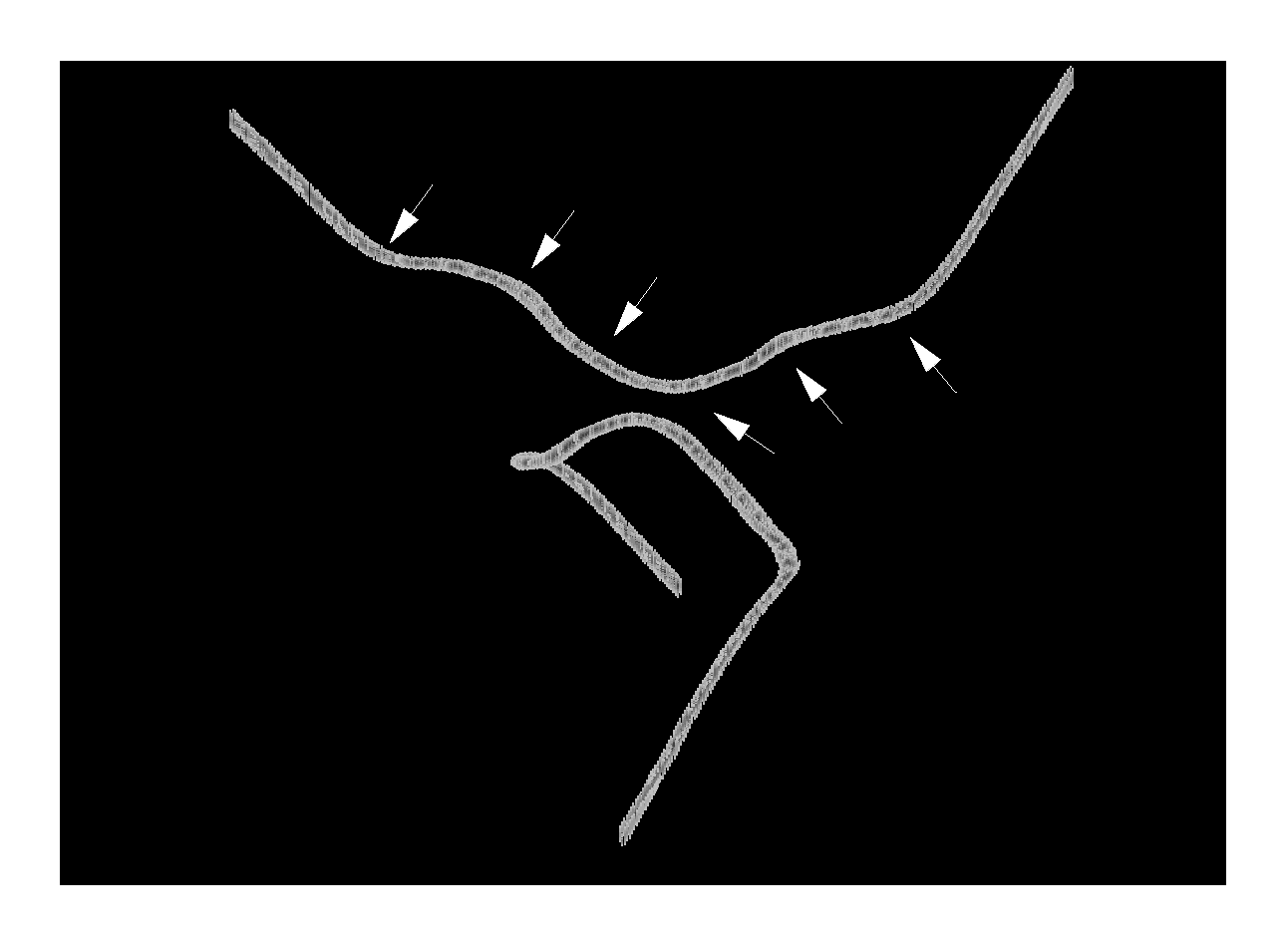} \\
\end{tabular}
\caption{\label{triple_panels} Isosurfaces of the scalar field with
  $|\phi|/\eta\leq 0.4$ for a collision with $(\beta = 32, \alpha =
  122.7, v = 0.88)$ showing a triple intercommutation. From left to
  right, up to down: snapshots at $t = 3,4.7,5,6.9,10,17,20,22$. Time is measured in inverse scalar masses (see text). At $t
  = 4.7$ the strings collide. Notice the distorsion around the point
  of collision (see also figure \ref{string_bending2D}). After the
  first intercommutation a loop emerges at $t = 6.9$. The loop catches
  up with the receding strings and intercommutes at the connection
  points ($t = 10$, second intercommutation), creating a highly curved
  central region in each string ($t = 17$). These move towards each
  other and produce a third intercommutation at $t = 20$. Two sets of
  three closely spaced, left- and right-moving kinks are visible on
  each of the strings in the last panel (indicated by arrows on the
  upper string).
}
\end{figure}
\begin{figure}
\includegraphics[height = 40mm,width=40mm]{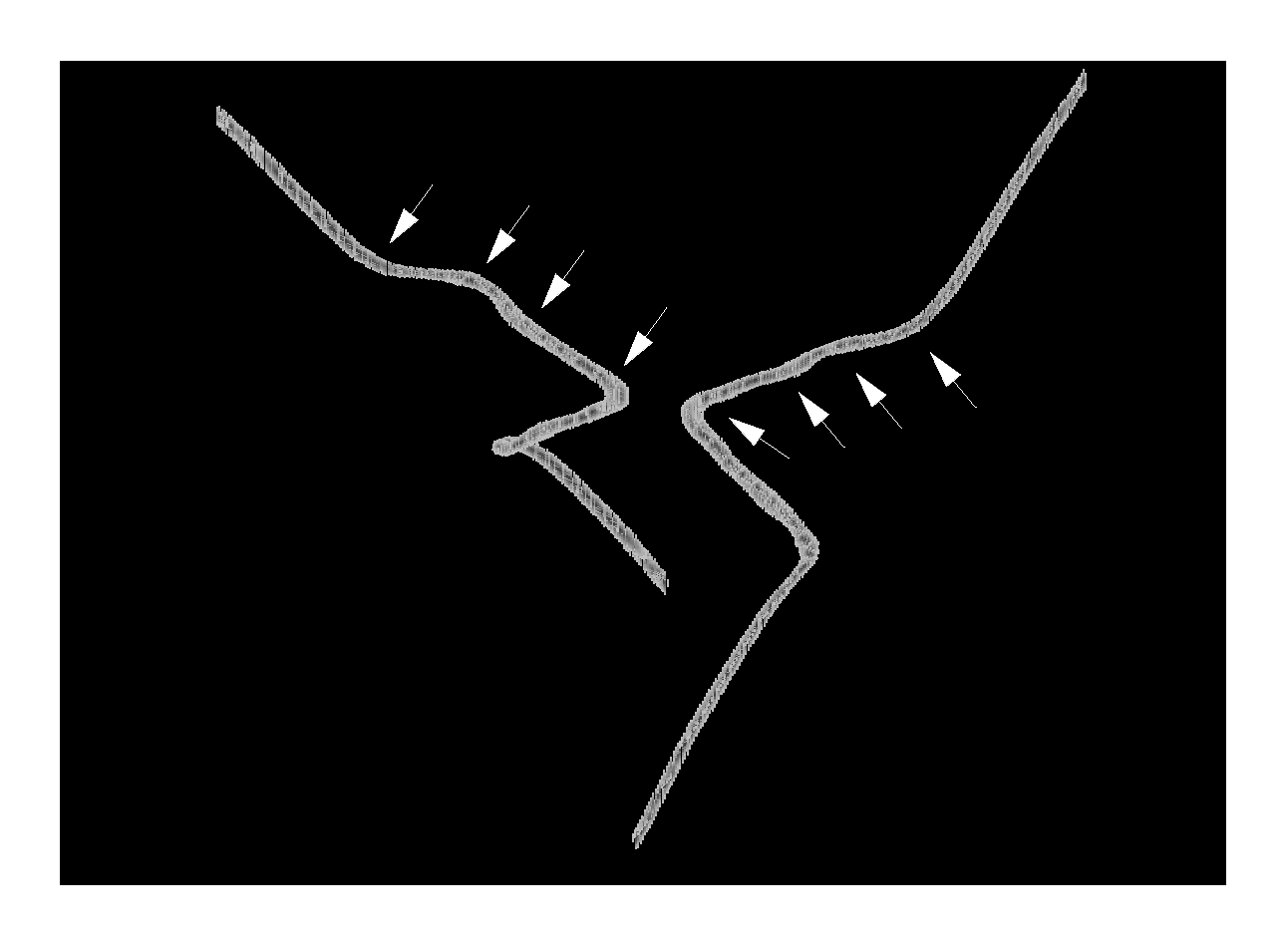}
\caption{After the third intercommutation there
  are two, almost anti-aligned, string segments close to each other
  (see the last panel of figure \ref{triple_panels}. If these are
  receding slowly, even a fourth intercommutation is possible. This
  figure shows the typical string configuration after the fourth
  intercommutation. The arrows show four closely spaced kinks moving
  up each string; the two corresponding down-moving trains are visible
  below the collision point. $(\beta = 16, \alpha = 122.7, v = 0.92)$,
  same isosurfaces.}
\label{quadruple_final}
\end{figure}
\begin{figure}
\begin{tabular}{cc}
\includegraphics[height = 40mm,width=40mm]{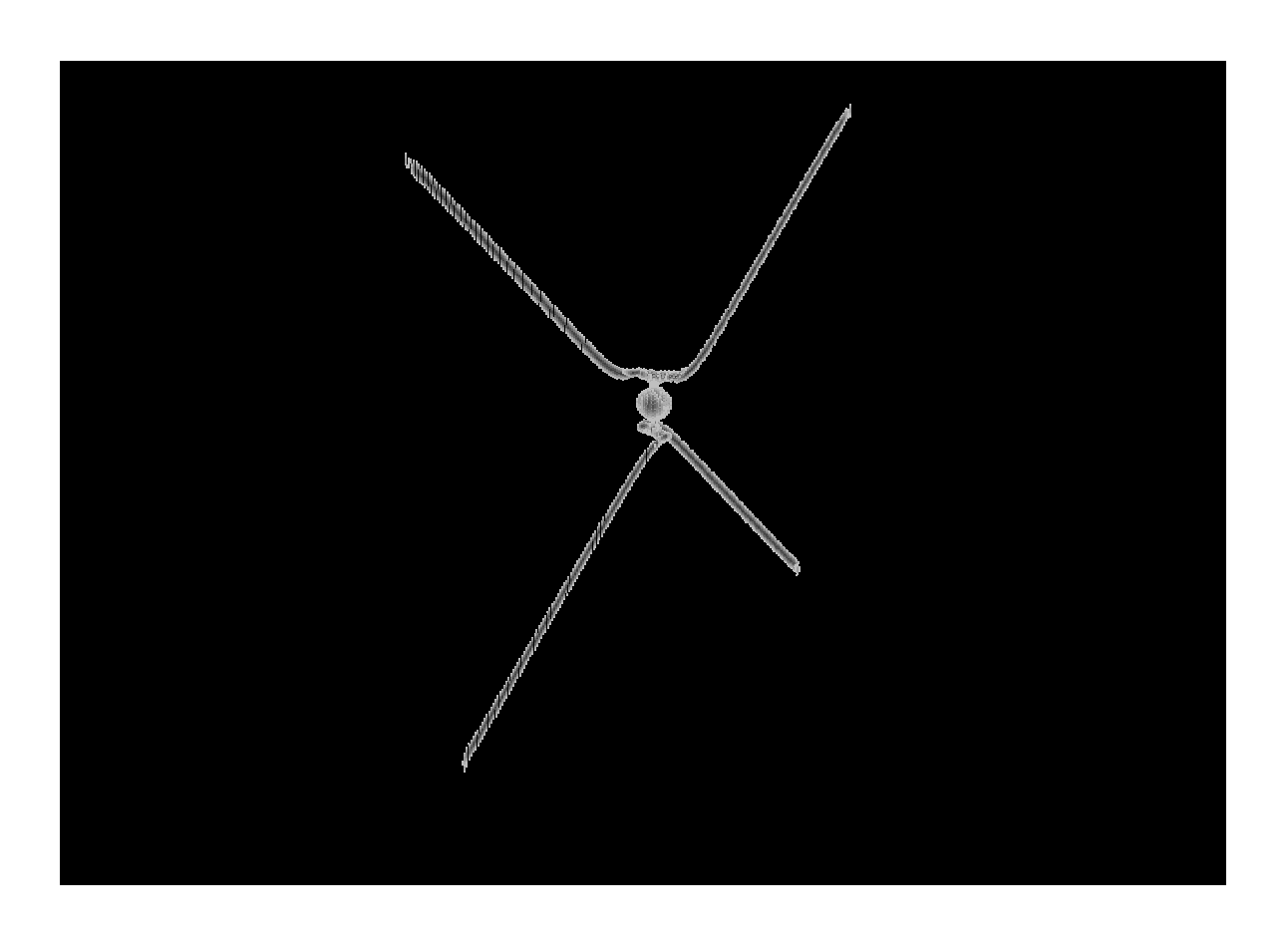} &
\includegraphics[height = 40mm,width=40mm]{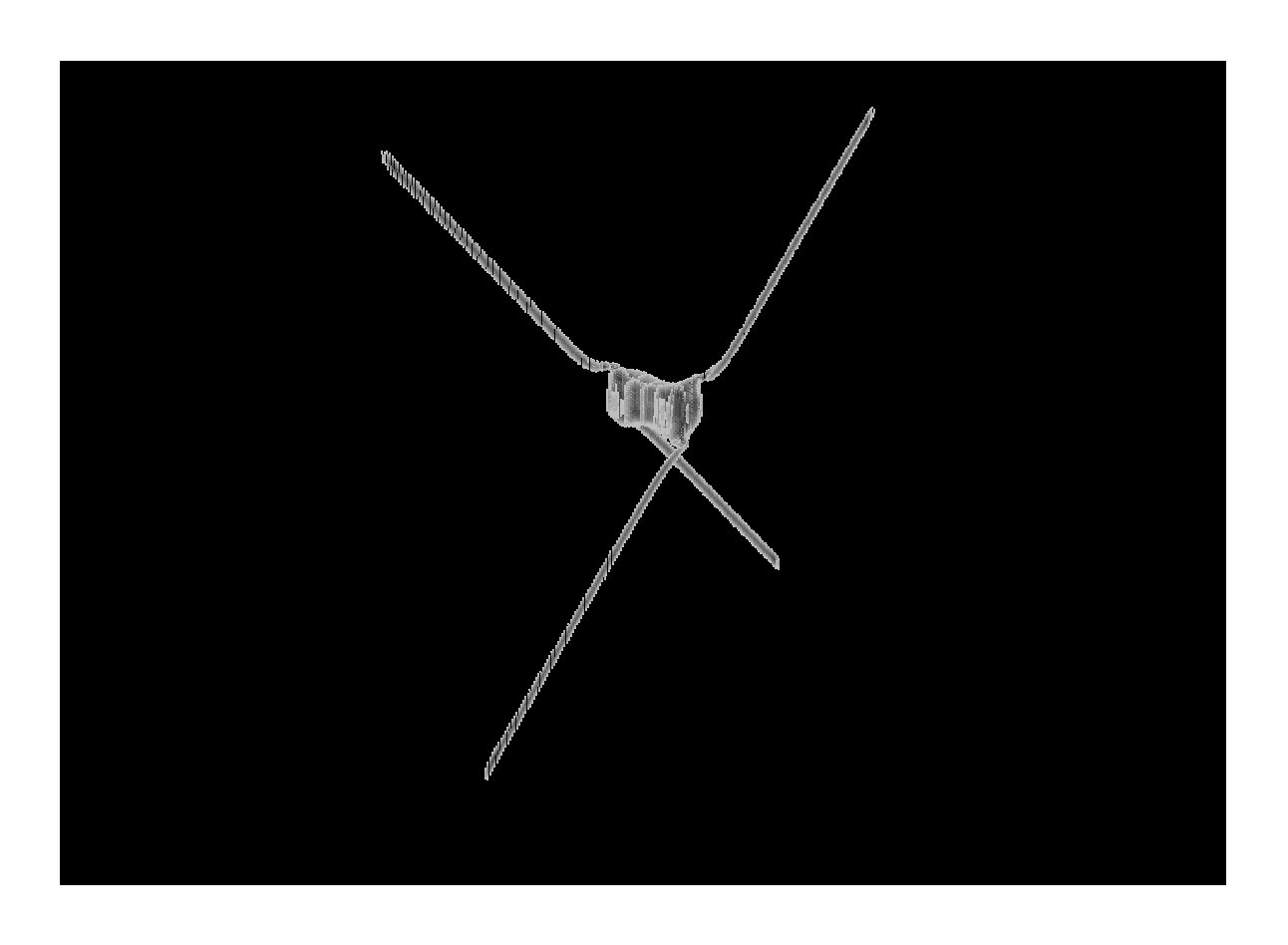} \\
\includegraphics[height = 40mm,width=40mm]{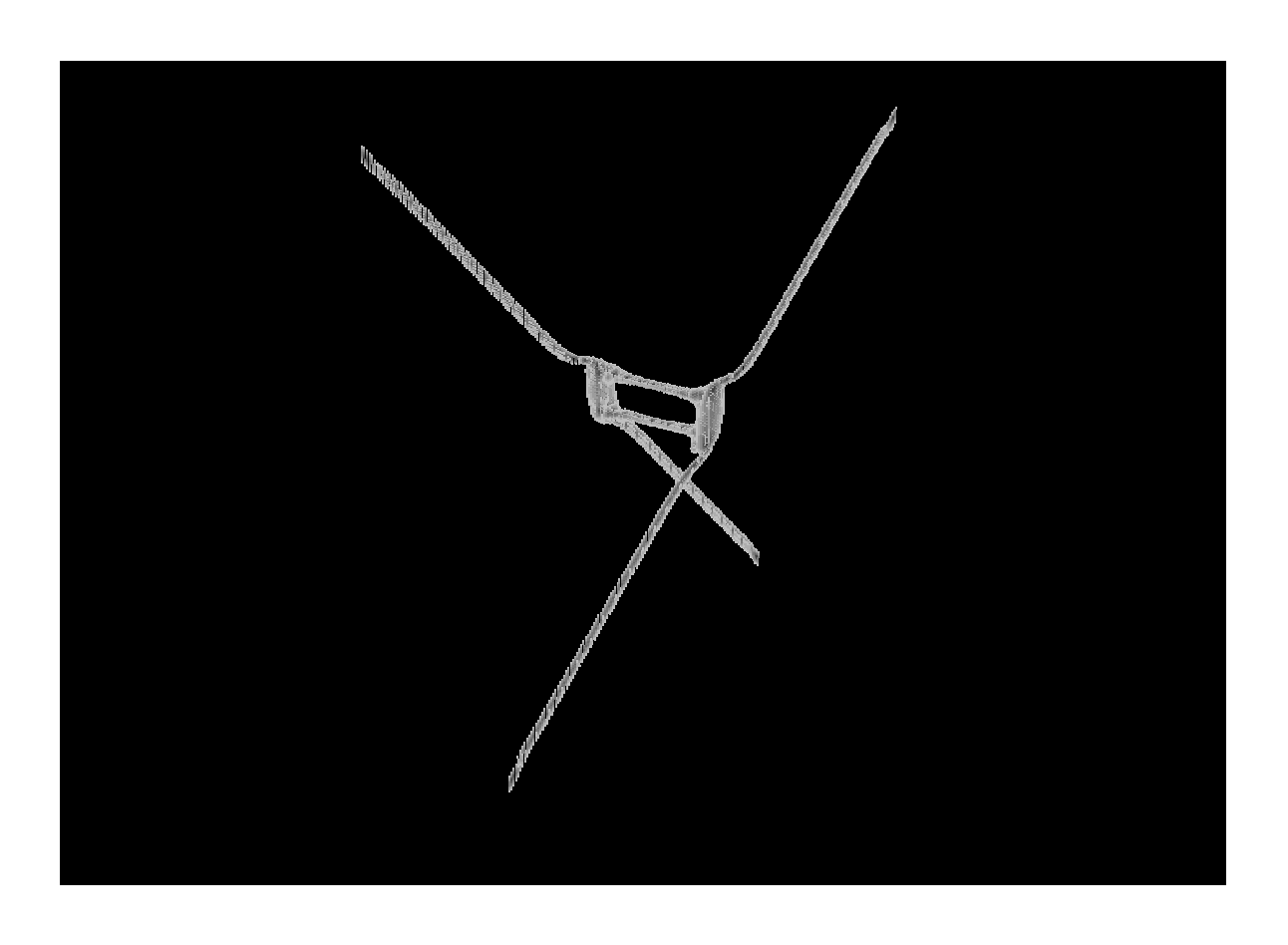} &
\includegraphics[height = 40mm,width=40mm]{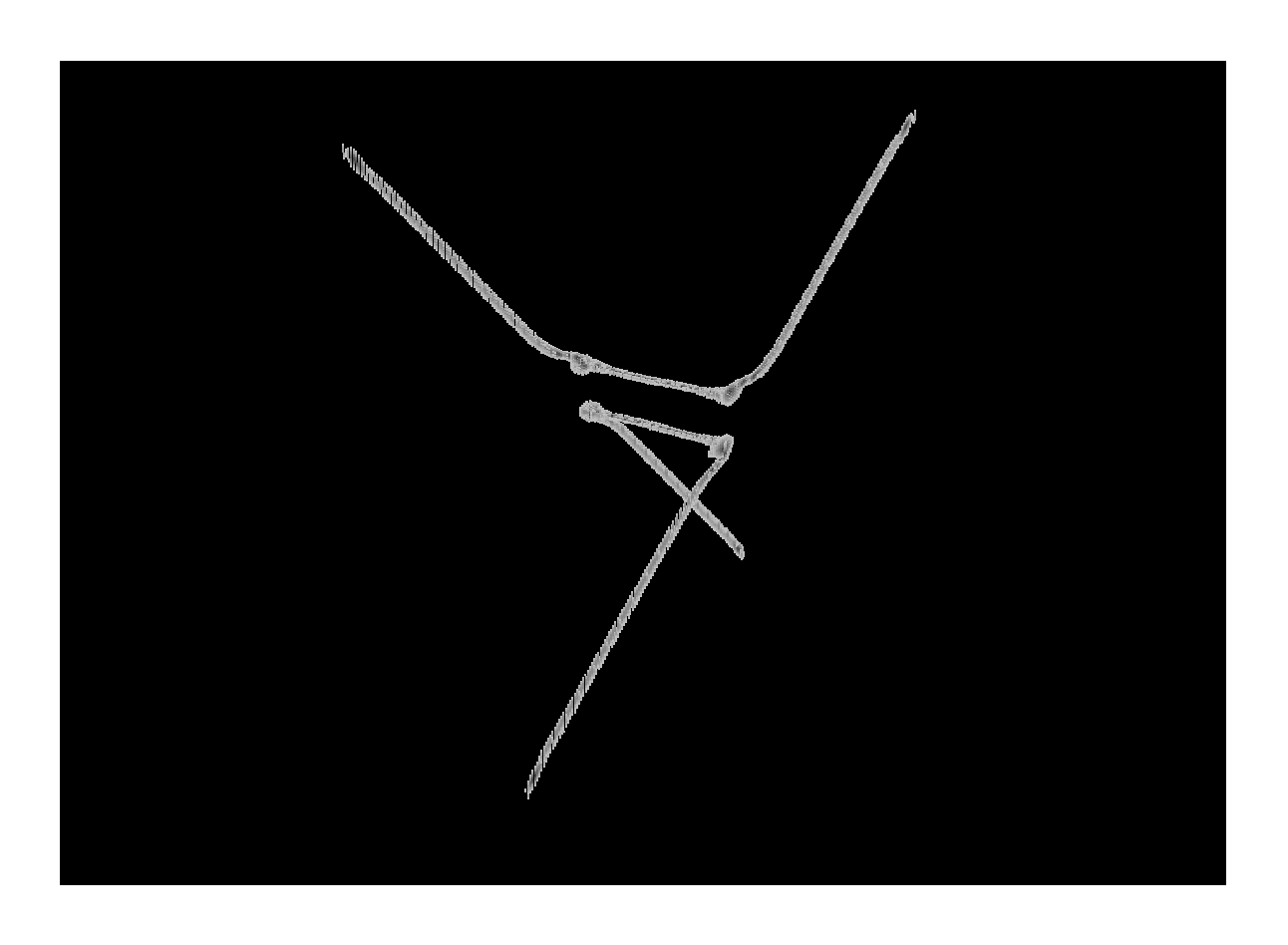} \\
\end{tabular}
\caption{After the first intercommutation a radiation blob
  emerges. The blob catches up with the bridge and is
  absorbed. However, before it is absorbed, a loop seems to be
  formed. This ``loop'' is not topological, it breaks resulting in
  "Dracula's teeth". $(\beta = 4, \alpha = 122.7, v = 0.98)$. The
  interaction between the blob and the strings slows them down,
  facilitating a second reconnection, but --unlike in figure
  \ref{triple_panels}-- the blob by itself cannot mediate this second
  reconnection. From left to right, up to down: snapshots at t = 2,
  4.5, 6.5, 8. Note the antialignment of the bridging segments between
  the strings, as in figure \ref{start_config}.}
\label{blob_panels}
\end{figure}

The images of the intercommutation process in figures
\ref{triple_panels}, \ref{quadruple_final} and \ref{blob_panels} show
isosurfaces of the scalar field with $|\phi| = 0.4$. At this value
only about $20\%$ of the potential energy is contained within them,
but it allows us to visualize the evolution of the Higgs field most
effectively. A tube twice the radius would contain about $60\%$ of the
energy (to be precise, a threshold of $|\phi| = 0.8$, which has twice
the thickness of the tubes shown, contains $62\% $ of the scalar
potential energy for $\beta = 16$, $57\%$ for $\beta = 32$
and $52\%$ for $\beta = 64$).\\

It is clear from these images that the
strings do not always intercommute once or twice, as previously
observed, but also three and four times for particular values of the
initial speed $v$ and angle $\alpha$. An odd number of reconnections
results in overall intercommutation of the strings, and an
even number in the strings effectively passing through, so we can
still speak of a threshold velocity for the strings passing
through. However, each reconnection creates small structure on the
strings in the form of a left- and a right-moving kink. In some cases
it is not easy to distinguish between one and three reconnections, or
between two and four reconnections, just by looking at the energy
isosurfaces in the intermediate state. But the resulting kinks are
clearly visible in the final state and can be counted; in
case of doubt we use this criterion (see fig. \ref{quadruple_final}).\\

Successive reconnections therefore lead to left- or right-moving
``kink trains'', groups of up to four closely spaced kinks (one for
each intercommutation). The inter-kink distance within these trains is
a few core widths, at the time of formation (see figs \ref{triple_panels} and
\ref{quadruple_final}).\\

A multiple intercommutation process for type II strings
unfolds as follows. After the collision in which the strings exchange
ends for the first time two things can happen:\\

$\bullet$ For $\beta \geq 16$, an expanding loop forms after a short
delay.  If $v > v_c$, the loop catches up with the two receding
strings and these reconnect again through the loop.  This creates a
highly curved central region in each string (sometimes, for the lower
collision angles, the loop is not as sharp as in figure
\ref{triple_panels} and the bridge is very pronounced, making the
intermediate state look more like a junction). The central regions
will move towards each other and in some situations mediate a third
reconnection (see fig. \ref{triple_panels}). After the third
reconnection there are two almost antialigned string segments, and if
they are receding sufficiently slowly a fourth reconnection is
possible. This is the largest number of intercommutations we have
seen, and only for $\beta = 16$. On the other hand, if $v < v_c$ the
second reconnection does not take place because the string loop does
not catch up with the strings. In this case it will contract again and
eventually decay into
radiation, sometimes after one or a few oscillations.\\

Triple intercommutations are quite generic for collision speeds and
angles on the boundary between the regions in parameter space where we
observe one and two reconnections. 
For $\beta = 64$ we see a few triple intercommutations in a window
around $v \sim 0.87, \ \alpha \sim 94^0$. The box is
already somewhat small and we expect a larger box with increased
dynamical range would show more multiple intercommutations, but this
remains to be confirmed.\\


$\bullet$ For $1< \beta \leq 8$ the energy isosurfaces look somewhat
similar, but they reveal a very different intermediate state. The
``loop'' in figure \ref{blob_panels} is just a blob of radiation with
no topological features: the (covariant) phase of the Higgs field
around the ``vortex'' that makes the loop shows no net winding around
the loop, $\int_0^{2\pi} (\partial_\theta \phi - i eA_\theta \phi)
d\theta = 0$. This is clearly visible in the third timestep, where the
``loop'' meets the string bridges, breaks and is absorbed -- a real
loop of string would not be able to break if it carried a net
winding. In this case the maximum number of reconnections is two.
The blob slows down the receding strings (thereby lowering the
critical velocity) and facilitates the second reconnection. However,
whether or not the second intercomutation takes place is still
determined by the string bridge (see fig. \ref{start_config}). We
therefore see, as expected, a good agreement between the data and
eqs. \ref{w_crit}-\ref{delta_crit} in fig. \ref{vcrit_panels} for
$\beta = 4$ and $8$.\\

We now turn to the simulations of the 2D head--on collision of a
vortex and an antivortex at ultrarelativistic speeds. The parameters
($\beta$, $v$) of the simulations we performed are listed in table
\ref{2D_simulations}.
While we confirm the general picture of ref. \cite{MyersRebbiStrilka},
we have slightly different results for the state after the collision: \\

\begin{figure}
\begin{tabular}{cc}
\includegraphics[height=40mm,width=40mm]{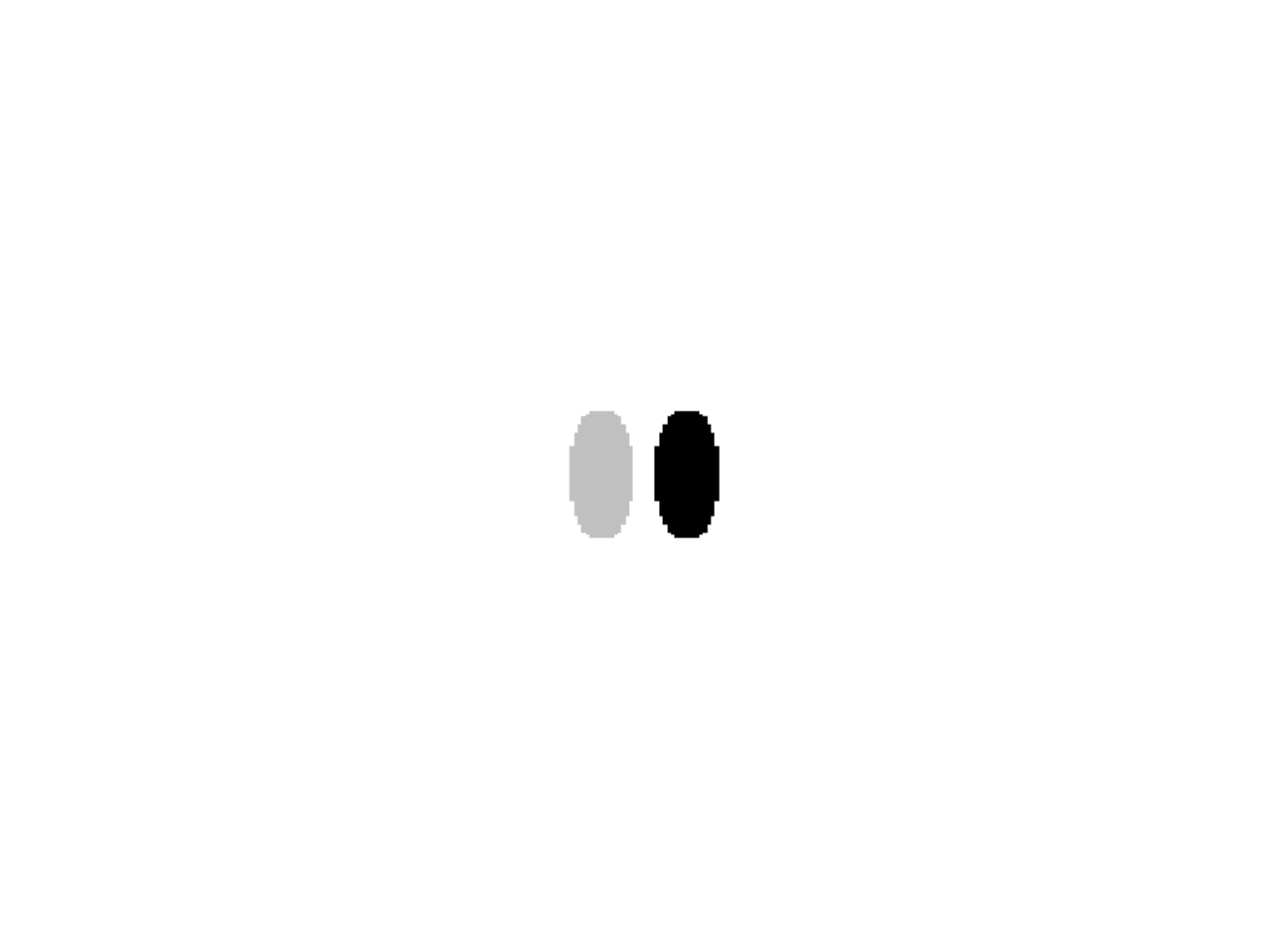}&
\includegraphics[height=40mm,width=40mm]{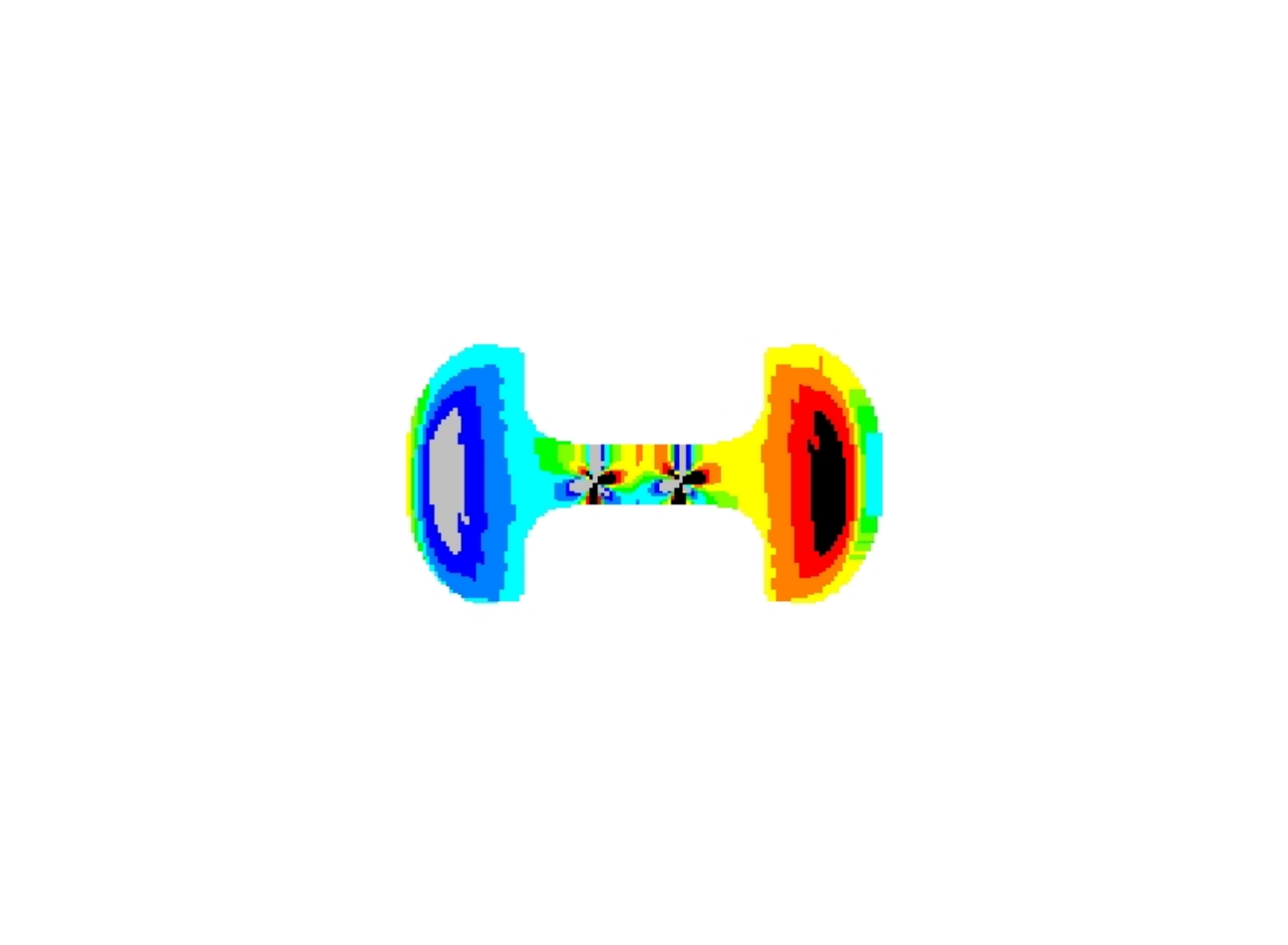}\\
\includegraphics[height=40mm,width=40mm]{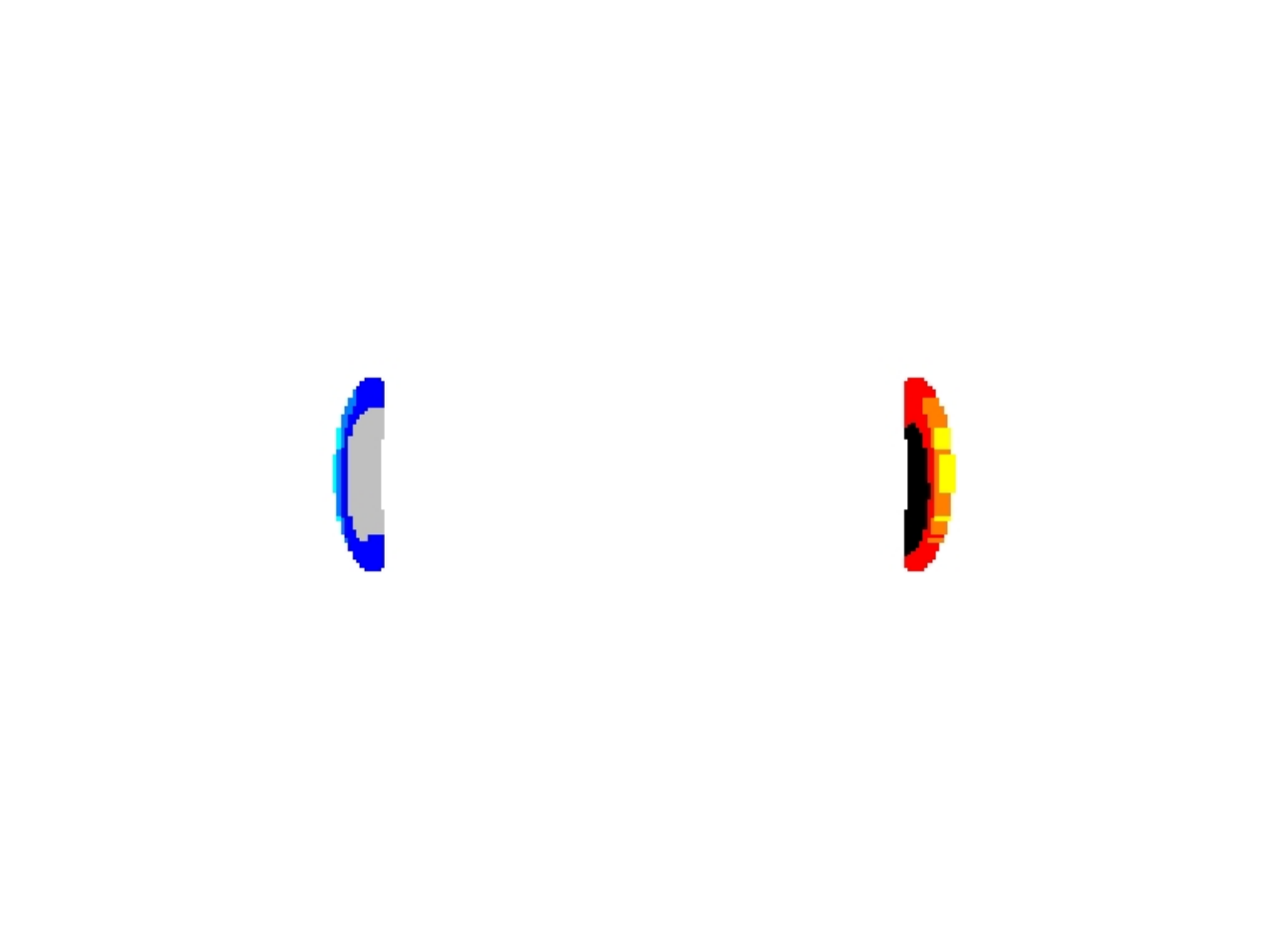}&
\includegraphics[height=40mm,width=40mm]{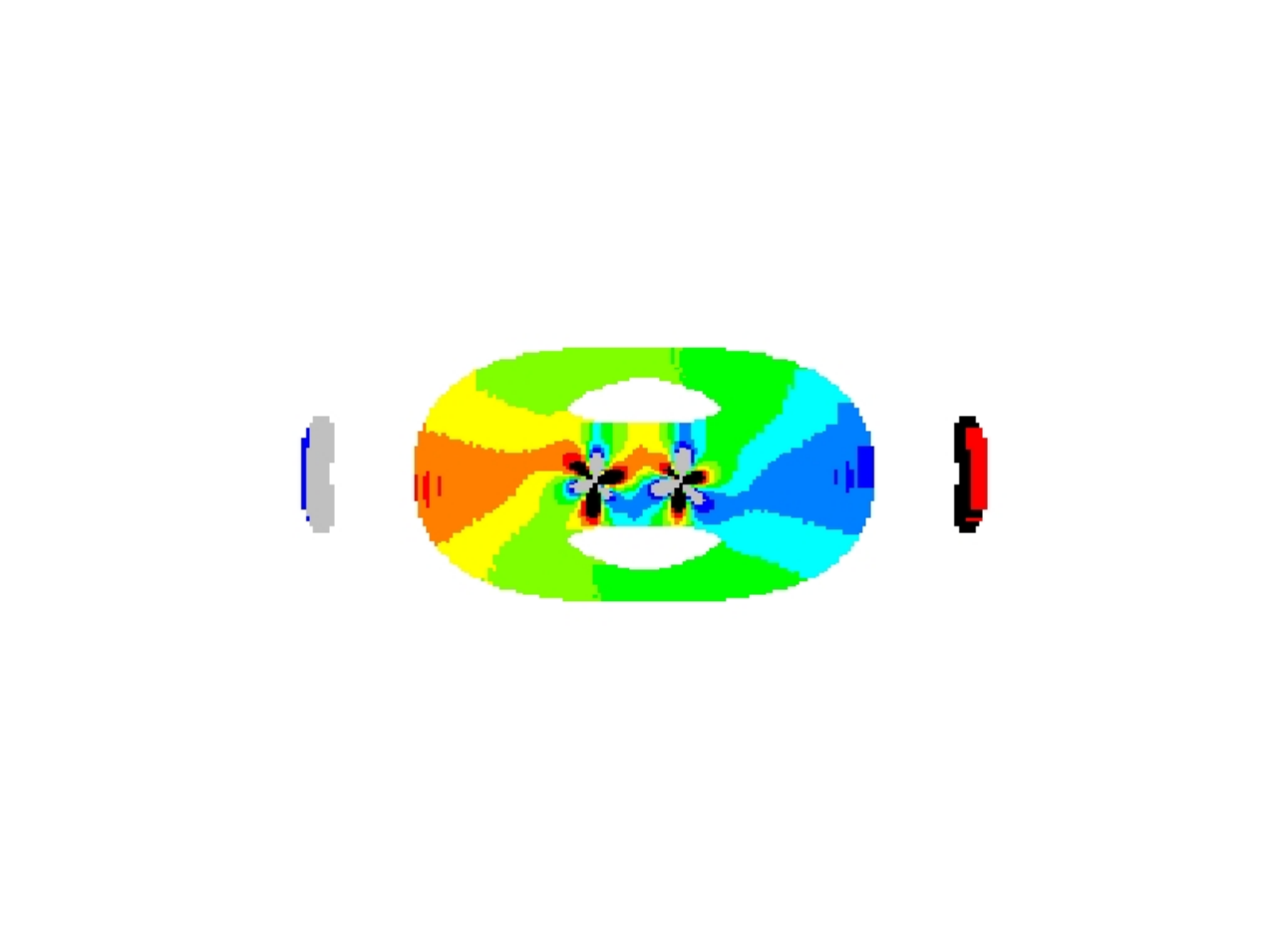}\\
\includegraphics[height=40mm,width=40mm]{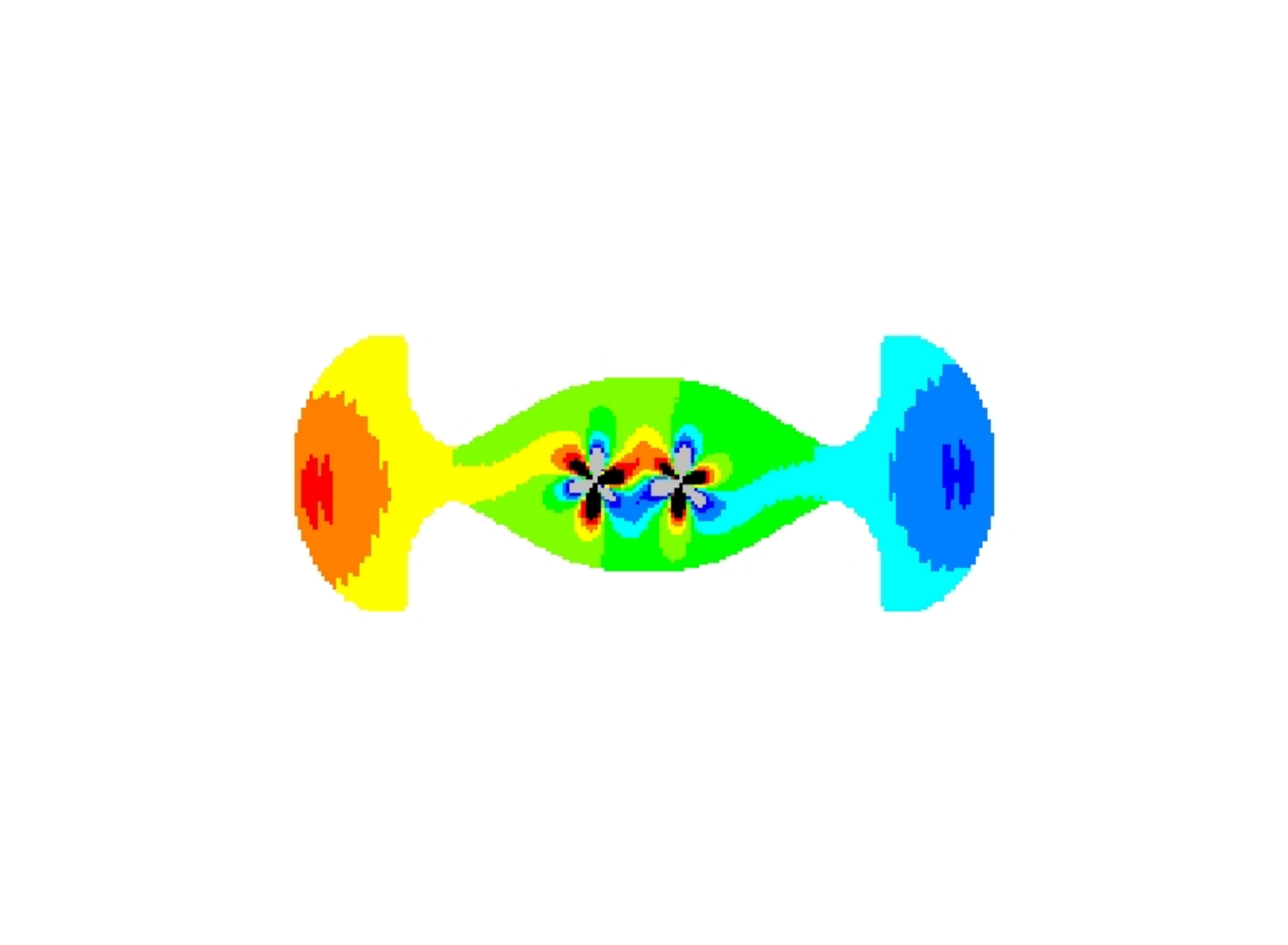}&
\includegraphics[height=40mm,width=40mm]{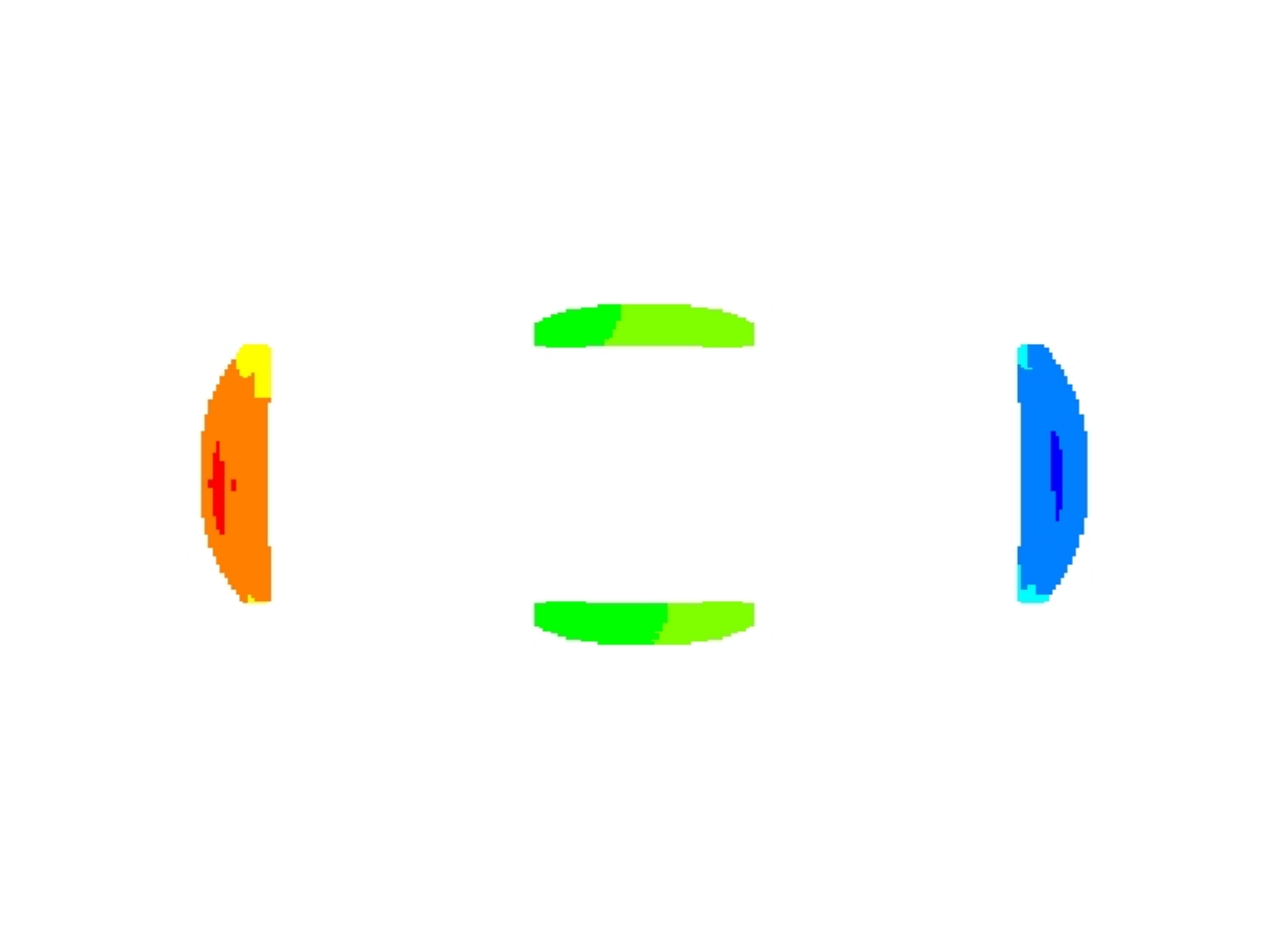}\\
\includegraphics[height=40mm,width=40mm]{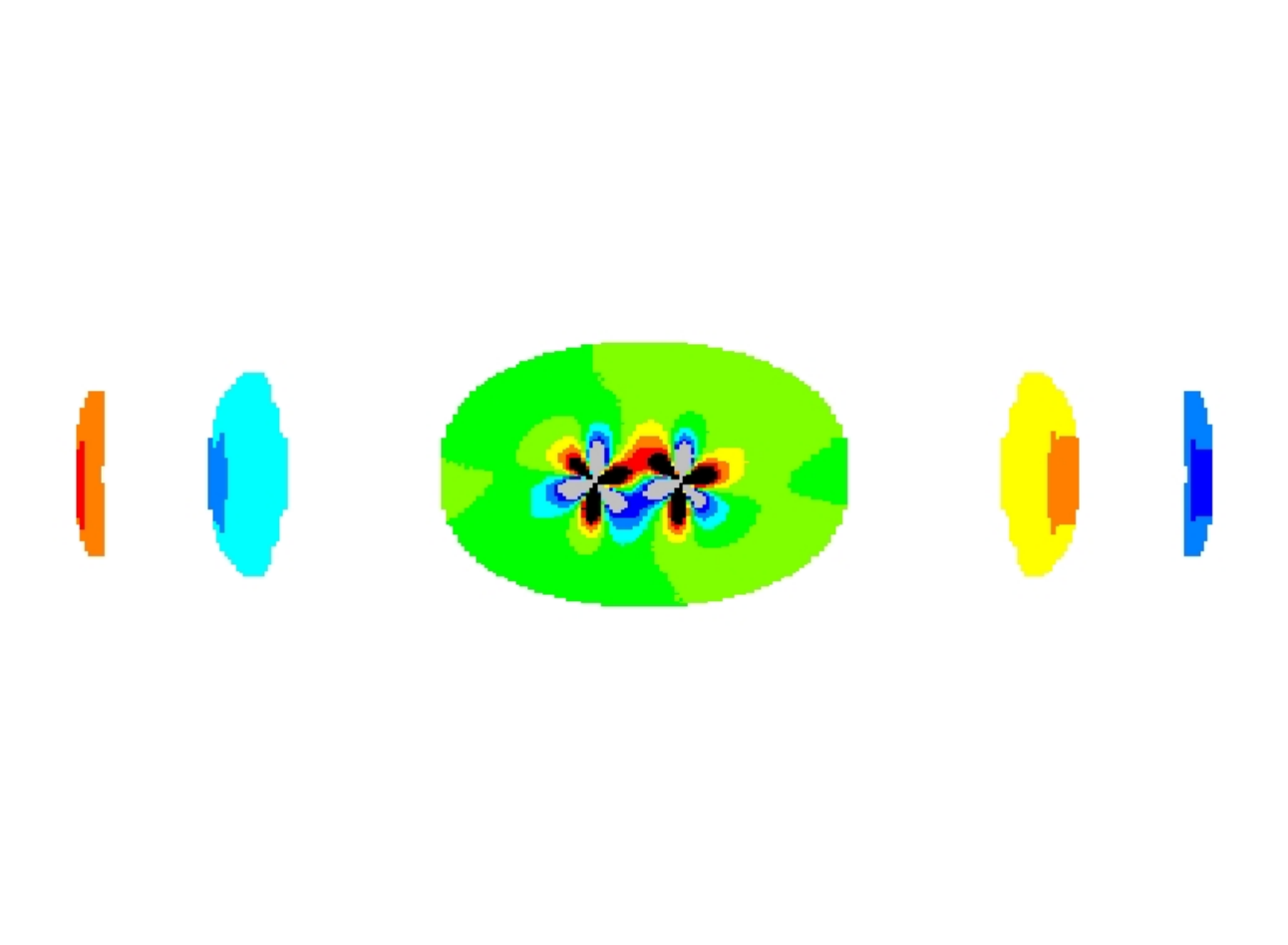}&
\includegraphics[height=40mm,width=40mm]{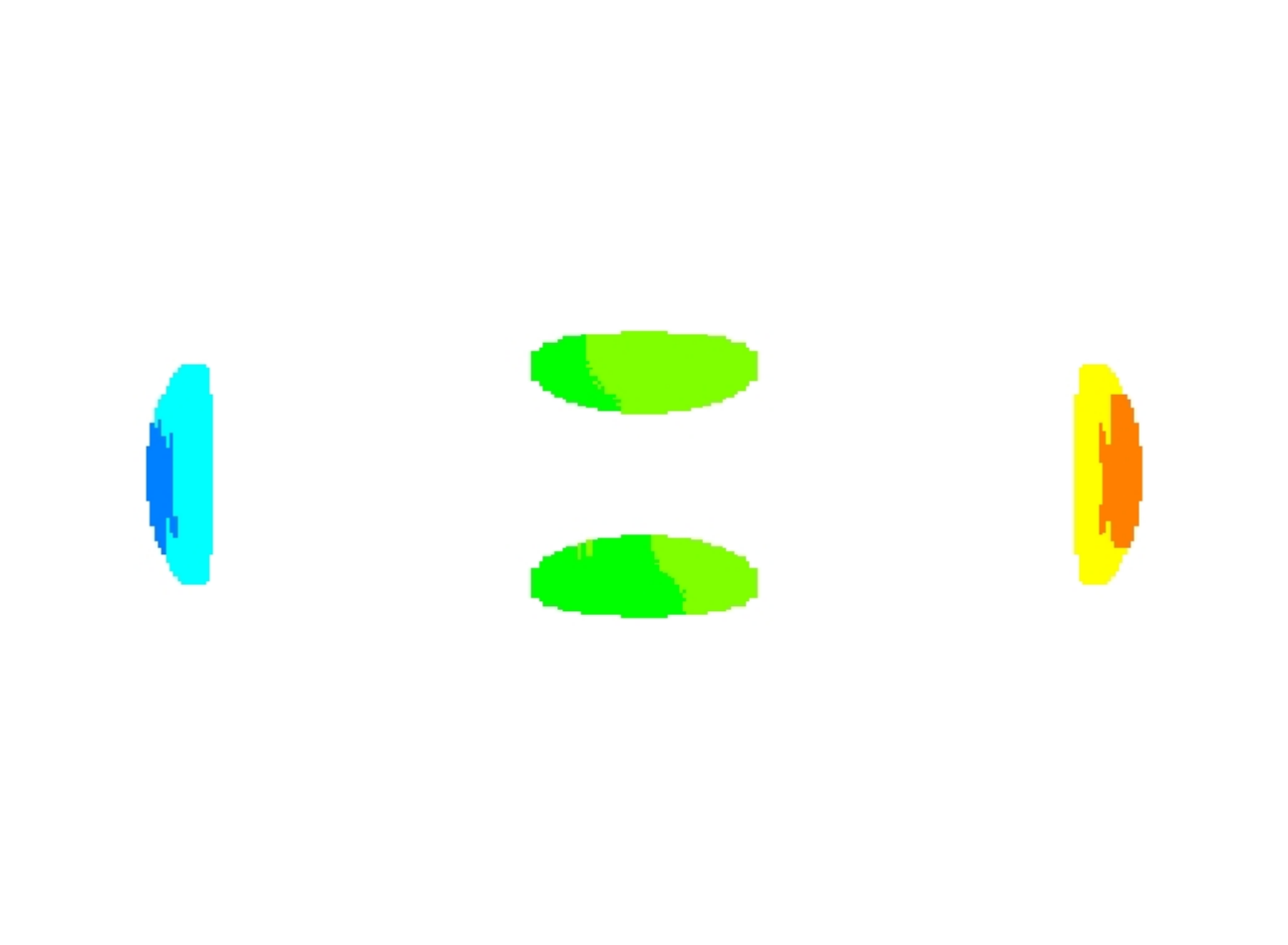}\\
\end{tabular}
\caption{A typical 2D head-on collision of a vortex and
  antivortex in the regime where where the outcome is a radiating
  bound state ($\beta$ below $6.2$). The figure shows
  the magnetic field of a vortex and antivortex collision with $\beta
  = 4$ and $v = 0.95$. All other parameters and timescales are like in
  the previous 3D simulations. The Grey to Blue colorscale indicates
  the magnetic field pointing out of the page, while Black to Red
  indicates the magnetic field points into the page. Black (grey)
  indicate points whose magnetic field is $\geq 20$\% of the maximum
  $B$ value, which is attained at the core of the incoming vortex
  (antivortex). The intermediate colours red/orange/yellow/light green
  (dark blue/blue/light blue/dark green) are in decreasing steps of
  5\%.  From left to right and top to bottom the panels show snapshots
  at t = 0,7,9,10,12,14,17 and 18. First, we see the vortex and
  antivortex at t=0. Second, we see a "pair" reform (t=7), but they
  are actually two blobs of radiation. In the third figure (t=9) we
  see a configuration which could be mistaken for a back-to-back
  reemergence. However, we then see a second blob forming (t=10), and
  the first pair fizzling out. The magnetic field oscillates around
  zero. At t=12 the second "pair" (blob) forms and some radiation is
  exchanged. By this time, the magnetic field of the first "pair" has
  already switched sign. This proves that the first "pair" had no
  topology. In the sixth panel (t=14) polarity is opposite to that in
  the second panel, but there is also radiation is the perpendicular
  direction. At t=17 a third blob forms and decays into radiation
  (t=18). This process continues for the dynamical range of the
  simulation or until all energy is radiated away.}
\label{2D_panels}
\end{figure}

$\bullet$ For $\beta \leq 6.2$, the emerging vortex and antivortex
settle in a bound, oscillating state, which completely decays into
radiation for all speeds between $0.9\leq v \leq 0.98$. This behaviour
extends to the $\beta <4$ regime studied by Myers et
al. \cite{MyersRebbiStrilka}, who interpreted the outcome as the
reemergence, back-to-back, of the original pair. We see no evidence of
this reemergence. A typical configuration after the
collision is shown in figure \ref{2D_panels}. Although some timesteps
could be mistaken for back-to-back reemergence of the
vortex--antivortex pair, subsequent evolution makes it clear it is only
localized radiation.\\

Note also that, since the vortex-antivortex pair never reforms after
the collision, the expectation of back-to-back reemergence, suggested
in \cite{srivastava} by analogy with the global vortex case, also does
not apply. \\

$\bullet$ For $\beta \geq 6.4$ and high collision speeds the pair reemerges as
if they passed through, and the critical velocity above which the
vortex-antivortex pair passes through each other goes down with
increasing $\beta$, from around $v = 0.98$ for $\beta = 6.4$ to
($\beta$, $v$): (6.6, 0.98), (6.8, 0.98), (7, 0.95), (8, 0.95) and
(32, $<$ 0.7). This agrees with ref.\cite{MyersRebbiStrilka}.

\begin{widetext}
\begin{tabular}{@{}l@{}@{}l@{}}
\includegraphics[width=86mm]{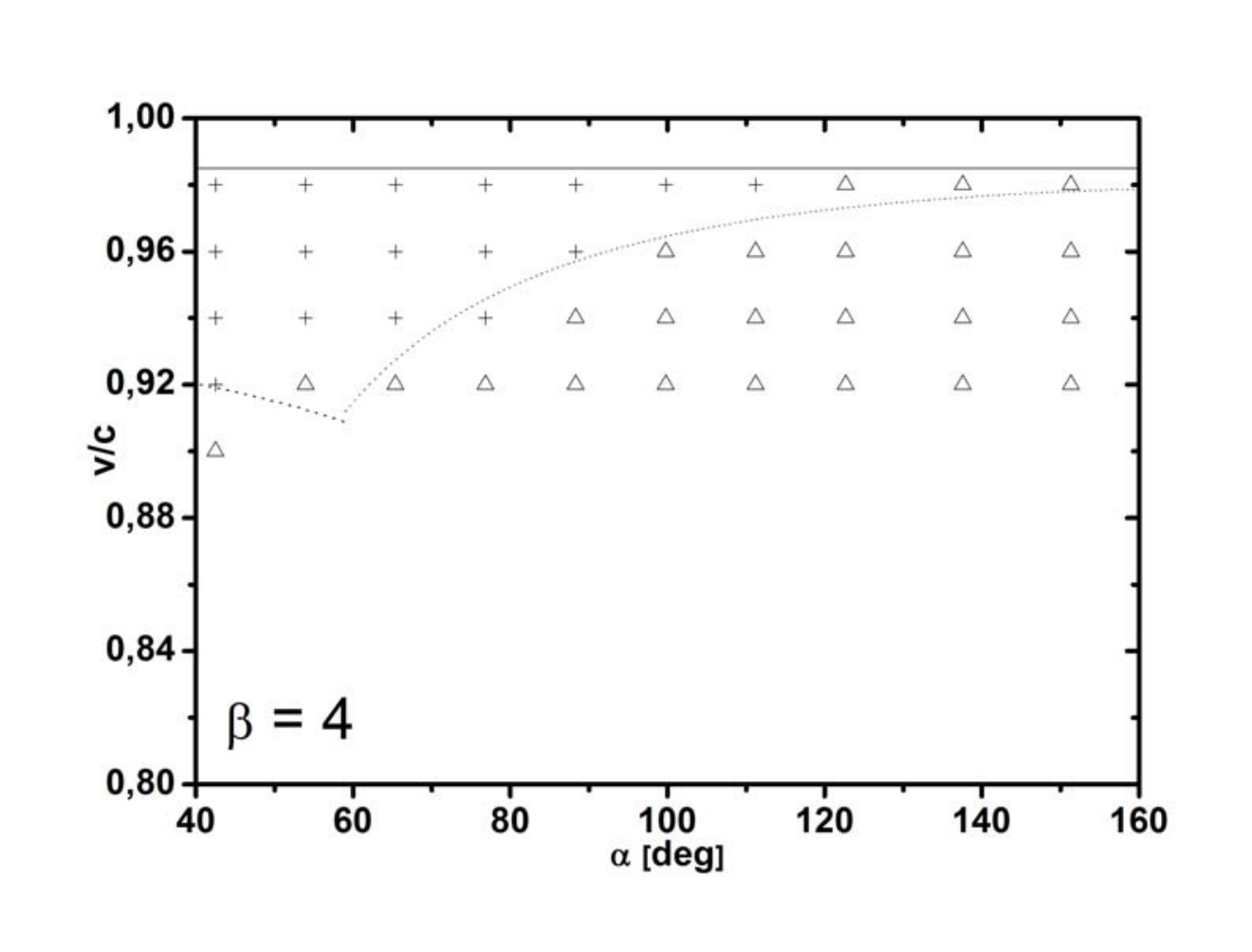}&
\includegraphics[width=86mm]{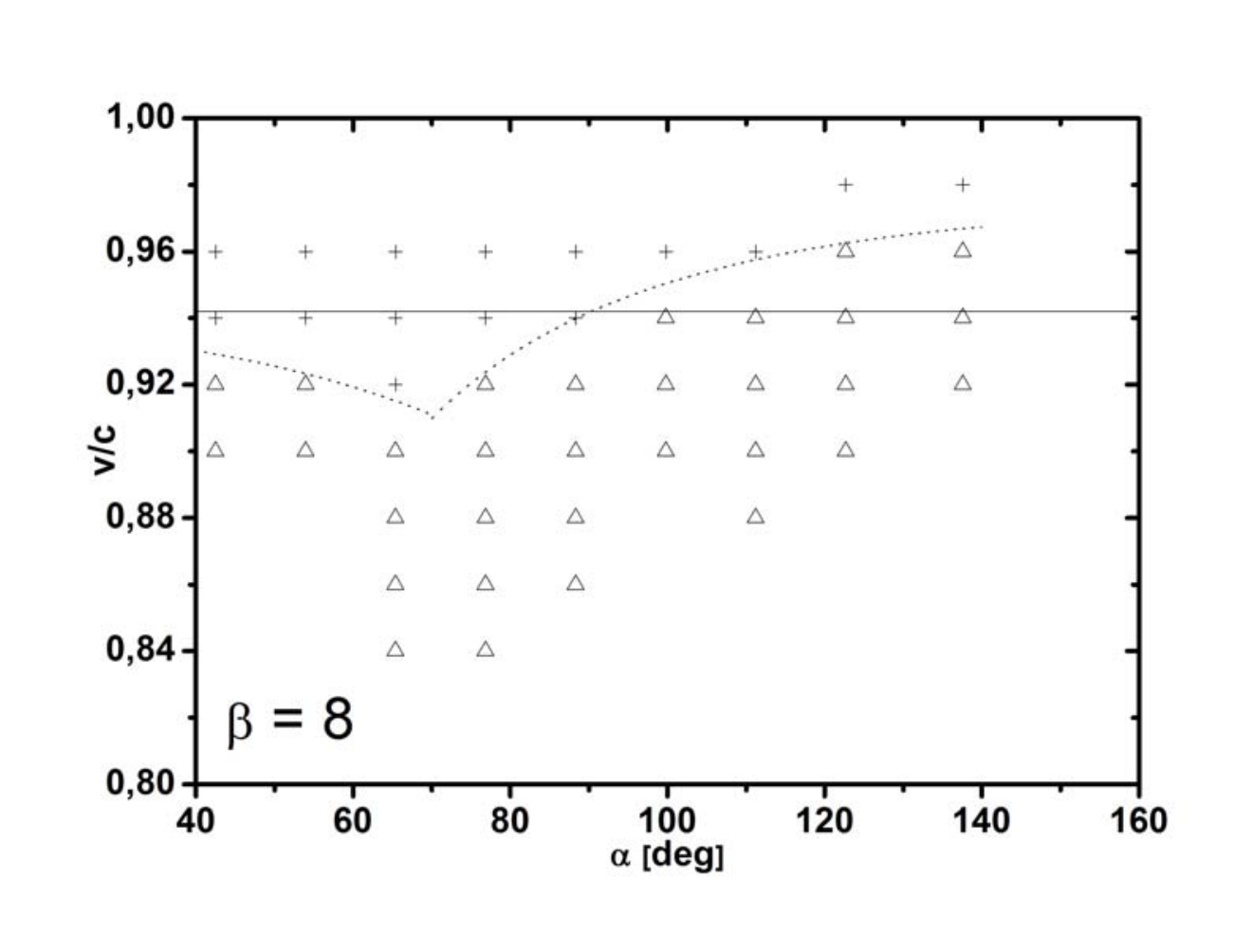}\\
\includegraphics[width=86mm]{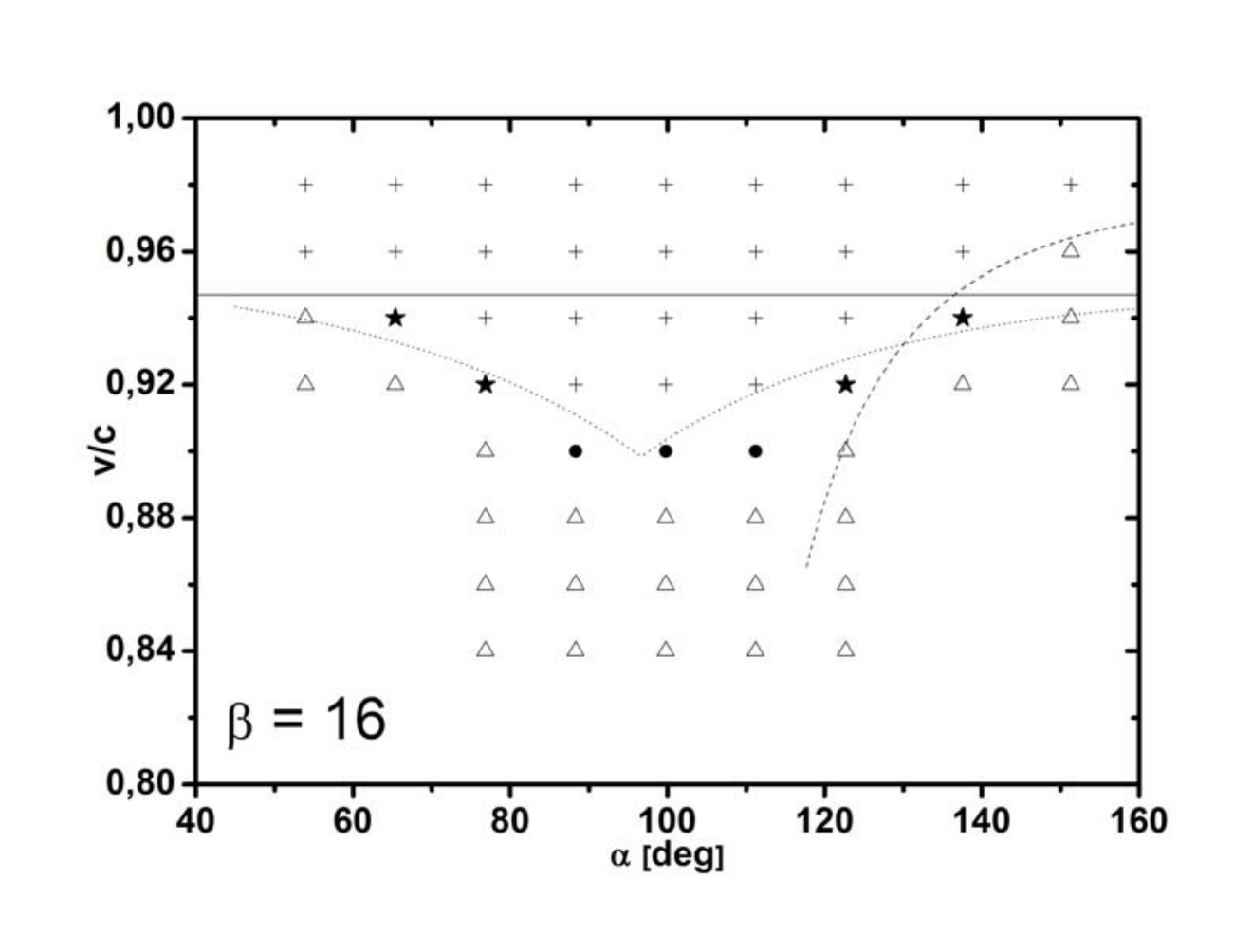}&
\includegraphics[width=86mm]{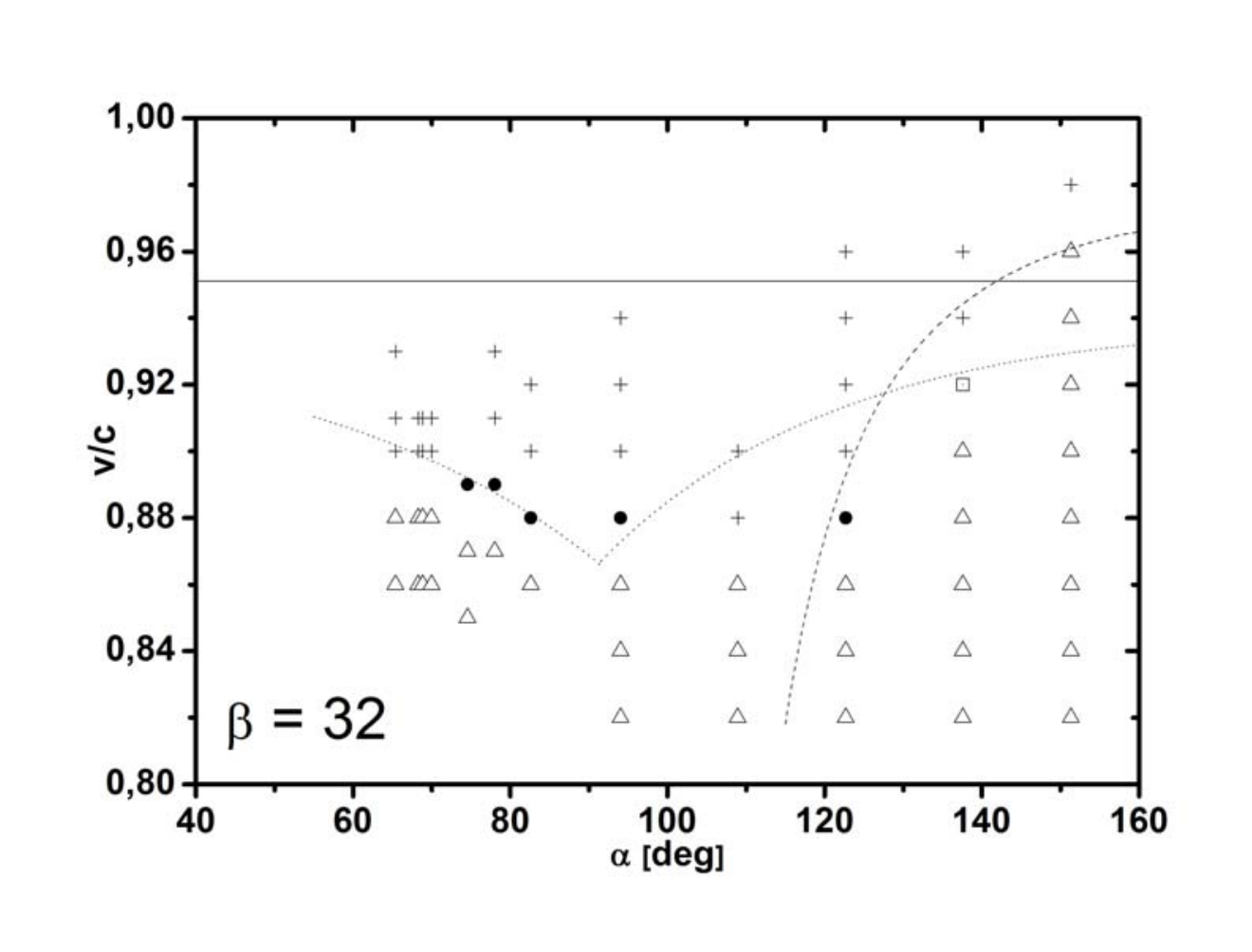}\\
\includegraphics[width=86mm]{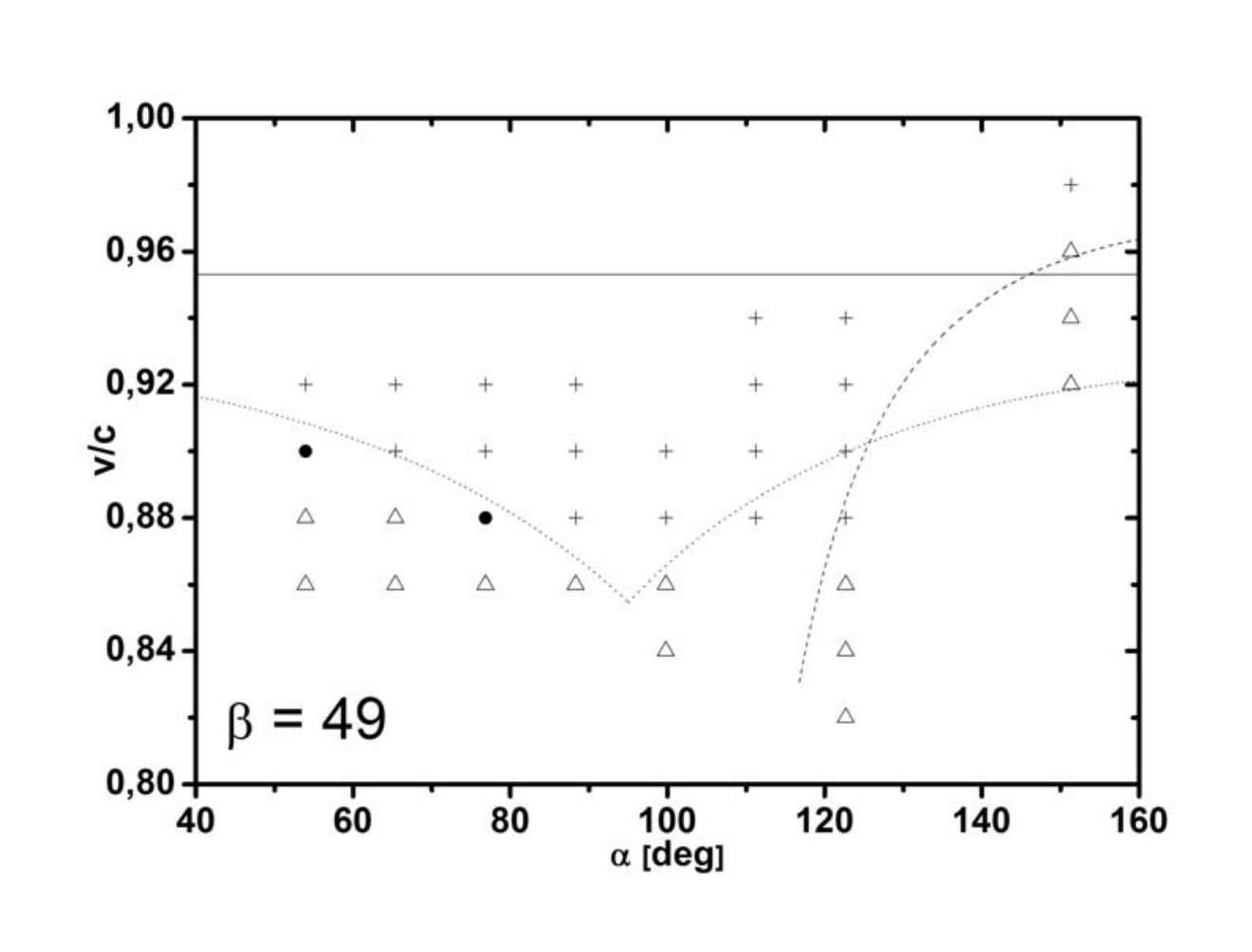}&
\includegraphics[width=86mm]{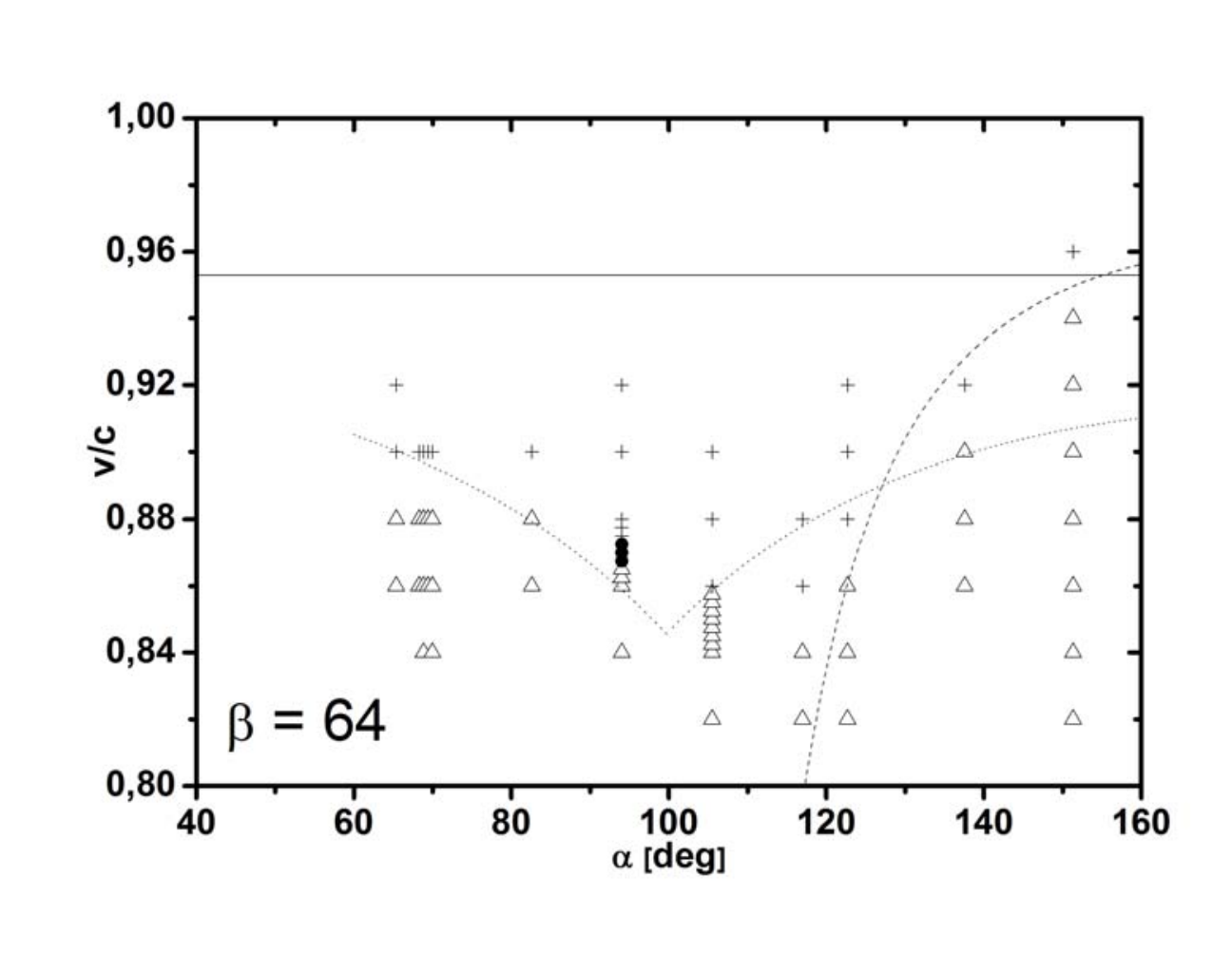}
\end{tabular}
\begin{figure}[!h]
  \caption{\label{vcrit_panels}
The number of intercommutations for a range
    of collision velocities and angles for $\beta = 4$, $8$, $16$, $32$, $49$
    and $64$. The symbols $\triangle$, $+$,$\bullet$ and $\star$ stand for
    1,2,3 or 4 intercommutations respectively.  The dotted lines show
    a two-parameter fit to simulations with collision angles below $150^0$, based on the Nambu--Goto approximation\cite{Putter}
    The dashed line is a one-parameter fit adapted to the deep type-II
    regime at high collision angle (see text for explanation and fit
    parameters) .  Simulations above the horizontal line resolve the
    scalar core size by less than three lattice points and are
    therefore less reliable. The point at ($\beta = 32, \alpha =
    137.6, v =0.92$), indicated by a square $\boxdot$, is inconclusive
    because the intercommutation happens just beyond the dynamical
    range. The $\beta = 4$ simulation has lattice spacing $a = 0.1$,
    all others have $a = 0.2$}
\end{figure}
\end{widetext}

\begin{table}
\begin{center}
\begin{tabular}{|c|r@{\,}l|}
$\beta$ & \multicolumn{2}{|c|}{$v$} \\
\hline
1 & 0.8, 0.85,&0.9, 0.95\\
1.01 & 0.8, 0.85,&0.9, 0.95\\
2 & &0.9\\
3 & &0.9\\
4 & (0.5), (0.6), (0.7), 0.8, 0.85,&0.9, 0.95*, 0.98\\
4.1 & 0.85,&0.9, 0.95, 0.98\\
4.2 & &0.9, 0.95, 0.98\\
4.3 & &0.9, 0.95, 0.98\\
4.4 & &0.9, 0.95, 0.98\\
4.5 & &0.9\\
4.6 & &0.9\\
4.7 & &0.9\\
4.8 & &0.9, 0.95, 0.98\\
4.9 & &0.9\\
5 & &0.9, 0.95, 0.98\\
6 & &0.9, 0.95, 0.98\\
6.2 & 0.85,&0.9, 0.95, 0.98\\
6.4 & 0.85,&0.9, 0.95, {\bf 0.98}\\
6.6 & 0.85,&0.9, 0.95, {\bf 0.98}\\
6.8 & 0.85,&0.9, 0.95, {\bf 0.98}\\
7 & 0.85,&0.9, {\bf 0.95, 0.98}\\
8 & 0.85,&0.9, {\bf 0.95, 0.98}\\
32 & {\bf 0.7, 0.8, 0.85, 0.875,}&{\bf 0.9, 0.925, 0.95, 0.98}
\end{tabular}
\end{center}
\caption{The parameters ($\beta$,v) of the 2D simulations of vortex--antivortex collisions described in the text. The simulations in boldface are those where the vortex--antivortex pair reemerges as if they had passed through. In all other cases the pair annihilates into radiation, sometimes after forming a short-lived pulsating bound state. In the simulations in parentheses, the vav-pair annihilate directly into radiation. The case $\beta = 4, v = 0.95$, indicated by a star, is shown in fig. \ref{2D_panels}}
\label{2D_simulations}
\end{table}



\section*{Discussion}

Our results suggest some interesting differences between
the $\beta >> 1$ regime and the much more studied $\beta =1 $ regime
when it comes to the intercommutation behaviour and the resulting
small scale structure. Some of these differences can be traced back to
the core interactions, in particular the repulsion between parallel,
deep type-II strings, which in a 3D setting can distort the impact
angles and velocities of the strings. The angular distorsions can be
parametrized and could in principle be incorporated in numerical
simulations of cosmic string networks and analytic studies of small
scale structure.

\subsection*{Multiple reconnections and the nature of the
  intermediate state}

As reported in \cite{us}, we have observed a qualitative change in the
process determining the second intercommutation. For $1 \leq \beta
\leq 8$ we see the emergence of a blob of radiation after the first
intercommutation. A blob cannot cause a second intercommutation by
itself, it is absorbed when it reaches the strings. In this case,
whether or not the second intercommutation takes place is determined
by the geometry after the first intercommutation. This situation is
well described by eqs. \ref{w_crit}-\ref{delta_crit} as found in
\cite{Putter}. On the other hand, for $\beta \geq 16$ we see the
emergence of an expanding topological loop that mediates the second
intercommutation.  We compare this transition with the reemergence of
the vortex-antivortex pair in highly relativistic two-dimensional
collisions: as $\beta$ increases, the critical speed for 2D
vortex-antivortex reemergence goes down {\it below} the critical speed
for double reconnection in 3D (and even below the 3D universal bound
on the critical speed $v\sim 0.77$ \cite{us}, see next
subsection). Therefore, for sufficiently large $\beta$, reemergence of
the v-av pair is unavoidable. In 3D the reemergent vortex-antivortex
pair leads to a loop.  Thus, for $\beta \geq 16$, the outcome is
always the same, the loop always forms.  However, whether or not this
loop leads to a second reconnection depends on whether the loop
catches up with the receding strings, which is, in turn, determined by
the velocity and angle of the strings before the collision: for high
$v$ there is a second reconnection, for lower $v$ there is only one
reconnection. It is also in this loop--mediated regime, of which the
lower bound is between $\beta = 8$ and $\beta = 16$ according to our
simulations, where the multiple or higher order intercommutations take
place. By contrast, for low $\beta$, the critical speed in 2D for the
reemergence of the v-av pair is so high that we do not see the
emergence of a string loop in 3D. Instead we see what we describe as a
radiation blob, the 3D equivalent of the bound, oscillating state in
2D. An interesting open question is to identify more precisely the
value of $\beta$ (between 8 and 16) at which the transition between
these two regimes occurs in 3D.

\subsection*{The critical velocity for double reconnection: $\beta$ dependence, angular dependence and a universal lower bound}



The dependence of $v_c$ with the collision angle shows that, as we go
deeper into the type-II regime, core interactions are playing an
important role, especially at high collision angles. With
better resolution and more data points than in \cite{Putter} we
actually see a difference with the $\beta \approx 1$ behavior. The
best fits with eqs. \ref{w_crit}-\ref{delta_crit} do not quite agree
with the data for large $\beta$ (e.g. $\beta > 16$). These
fits underestimate the critical velocity for impact angles close to
antiparallel (or else do not properly account for impact angles below
$90^0$). Discarding data points with impact angles larger than $150^0$
leads to the fits shown in figure \ref{vcrit_panels} with the dotted
line. We found for $\beta = 4, 8$ the following parameters: $w_t =
0.202, 0.238$ and $\delta_t = 136.4^o, 139.3^o$.  For $\beta = 16, 32,
49$ and $64$ we found $w_t = 0.328, 0.357, 0.261, 0.408$, $\delta_t =
144.0^o, 136.1^o, 139.0^o, 135.7^o$ respectively.
(we note that for $\beta = 49$ the fit has fewer relevant data
points).\\

The deviation between the fits and the data at high collision angle
is, by itself, not so surprising since in the deep type-II regime two
things change: First, the appearance of the loop means the angle
between the string segments is not expected to be relevant. The second
intercommutation will occur if the loop touches the receding segments,
irrespective of their mutual angle. Second, the core interaction
affects the state before the collision, it produces a torque that
deforms the strings and tends to anti-align the colliding
segments. This means the true collision angle is actually larger than
the initial value.
The torque results from the attraction between vortex and antivortex and vortex-vortex repulsion in the orthogonal plane, which compete in the type-I regime but add up in type-II.\\

\begin{figure}
\includegraphics[width=50mm]{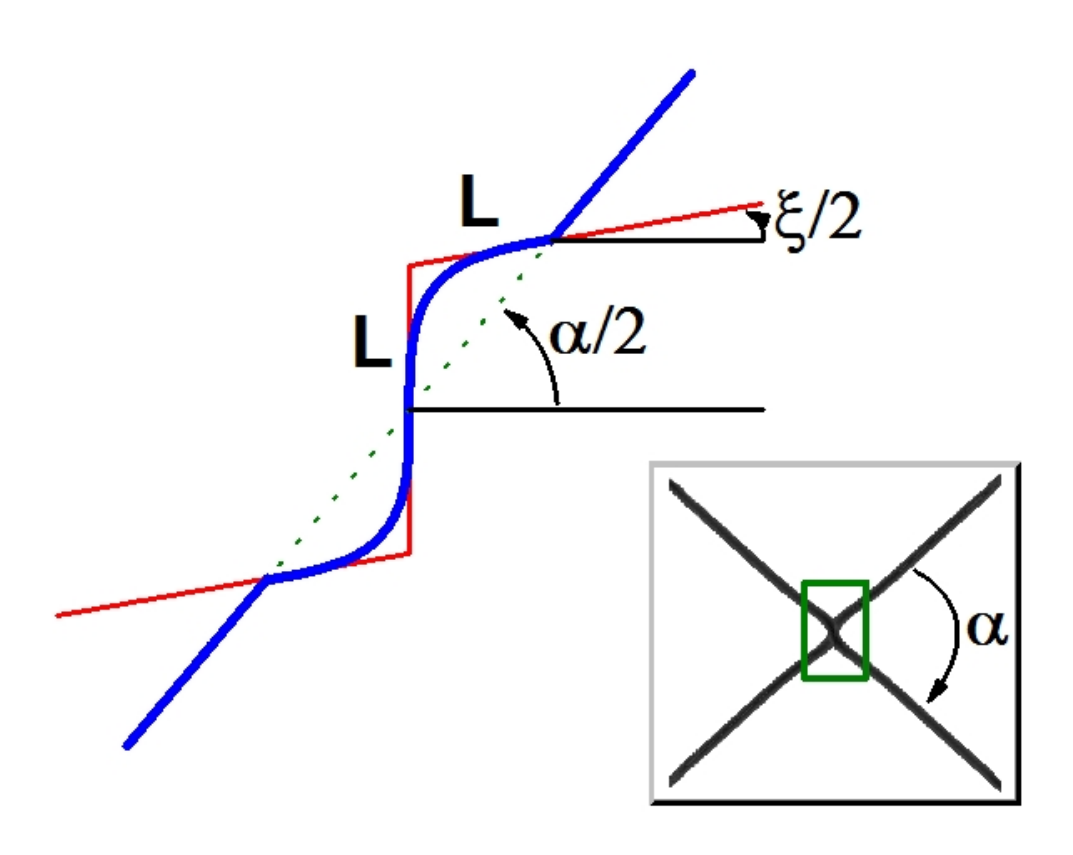}
\caption{The bending of the strings at the point of collision due to
  the intervortex potential, modelled in the figure as a sinusoidal
  perturbation. $\alpha$ is the initial collision angle when the
  strings are approaching but still far apart. The torque tends to
  anti-align the segments around the collision point (see the second
  panel of figure \ref{triple_panels}). Subsequently, a vertical
  segment of length $\sim 2L$ will annihilate from each string,
  leaving the receding strings as if they had collided with angle
  $\xi$. Note that, although the torque tends to increase the
  collision angle, the effective reconnection angle $\xi$ is actually
  {\it smaller} than the initial collision angle
  $\alpha$. \label{string_bending2D}}
\end{figure}

One might expect that a modification of the NG fit of \cite{Putter} to
account for this offset in collision angle should give a better fit in the deep type-II regime:
\begin{equation}
\label{w_crit_offset}
w = \sin [(\alpha - \alpha_0)/ 2] {1\over \gamma(v)}
\end{equation}
\begin{equation}
\label{delta_crit_offset}
\cos(\delta/2) = \frac{\cos [(\alpha - \alpha_0)/2]/(v\gamma(v))}{\sqrt{1+ (\cos [(\alpha - \alpha_0)/2]/(v\gamma(v)))^2}}
\end{equation}
with $\alpha_0$ negative. This expectation is not borne out by the
data. In fact, if anything, the data prefer a modification with a {\it
  smaller} angle (positive $\alpha_0$ -- see table \ref{offset_fit}). And this can be understood by
looking at figure \ref{string_bending2D}.  Although the actual
collision angle (the angle at the point of collision) is larger than
$\alpha$, the motion of the strings after the collision is determined
by an {\it effective} collision angle $\xi$ that is {\it smaller} than
$\alpha$.  This is because the antialignment causes a larger portion of
string to be annihilated so, after a quick transient, the strings look
as if they had collided with angle $\xi < \alpha$ and the bridge
segments (see fig \ref{start_config}) are further apart than they
would have been in the absence of a torque.

If the collision is sufficiently close to antiparallel, the relation
between the angle of approach, $\alpha$, and the effective collision
angle $\xi$ is universal (see fig. \ref{string_bending2D}): $\tan
{\alpha / 2} = {(1 + \sin {\xi / 2} ) / \cos {\xi / 2}}$, that is,
$\xi = 2 \alpha - \pi$, or $\alpha_0 = \pi - \alpha$, expected to be
valid for large $\alpha$.  .
This one-parameter fit is shown with a dashed line in figure
\ref{vcrit_panels}. The fit parameter is $w_{crit}$ and it has values
($\beta$,$w_{crit}$): (16, 0.233), (32, 0.243), (49, 0.251) and
(64, 0.275).\\


\begin{table}
\begin{center}
\begin{tabular}{|c|c|c|c|}
$\beta$ & $w_t$ & $\delta_t$ & $\alpha_0$\\
\hline
4 & 0.227 & 137.2 & -20.7\\
8 & 0.249 & 141.5 & -6.8\\
16 & 0.238 & 223.9 & 50.6\\
32 & 0.282 & 229.7 & 32.8\\
49 & 0.266 & 233.3 & 48.1\\
64 & 0.257 & 234.7 & 58.7
\end{tabular}
\end{center}
\caption{The best fit parameters of eqs. \ref{w_crit_offset} and \ref{delta_crit_offset}. A positive (negative) offset $\alpha_0$ indicates the actual collision angle is smaller (larger) than in the initial configuration. We see that $\alpha_0$ becomes positive for $\beta \geq 16$ and remains positive and relatively large.}
\label{offset_fit}
\end{table}


The offset in collision angle (see table \ref{offset_fit}) for $\beta
\geq 16$ indicates that the torque is strong even at very high
collision speeds, and suggests strong distortions for low speed
collisions.  This is potentially very interesting from the point of
view of the radiation coming from cosmic strings. Usually, the
radiation bursts from reconnection are subdominant to those from cusps
and kinks \cite{DamourVilenkin01,JacksonSiemens}.  But here reconnection
bursts are enhanced because longer segments of string are
annihilated. The amount of radiation produced by cosmic string
reconnection might therefore be somewhat larger in deep type II
collisions than would be expected from analytic, Nambu--Goto arguments.

Finally, our results appear to confirm the claim \cite{Putter} that
the critical speed goes down with increasing $\beta$, from $v_c \sim
0.96$ at $\beta = 1$ to $v_c \sim 0.86$ at $\beta = 64$, although this
reduction cannot go on indefinitely.  A crude (universal) lower bound
for $v_c$ follows from energy conservation \cite{us}: if the strings
anti-align locally and a portion $L$ of each string is annihilated,
the maximum energy available to the loop is $2L\gamma = 2 \pi R$, with
$R$ the maximum loop radius. If $R < L/2$ the second intercommutation
cannot take place, which happens for $v < \sqrt{1 - 4/\pi^2} \sim 0.77
$. These values of $v_c$ are to be contrasted with the average root
mean square velocity of the network, which has not yet been
investigated for type-II strings (known values range from $v_{rms} =
1/\sqrt 2 \sim 0.71$ for Nambu--Goto strings in flat space, to $~ 0.63
- 0.51$ in field theory simulations with $\beta = 1$ with cosmic
expansion \cite{MooreMartinsShellard,HindmarshStuckey}). So, double
intercommutations may be less rare than in type-I collisions, but they
are still rare events.\\

\section*{Summary and outlook}

We have investigated numerically the intercommutation of Abelian Higgs
strings in the deep type-II regime for selected values of $\beta
\equiv m_{scalar}^2/m_{gauge}^2 >> 1$ up to $\beta = 64$, the highest
value studied to date. Our study shows interesting differences with
the $\beta = 1$ behaviour and raises some puzzling questions.  New
effects arise due to the strong interactions of the string cores.
Multiple reconnections were already reported in \cite{us}, and also a
qualitative change in the intermediate state after the first
reconnection, with truly topological loops only appearing for $\beta
\geq 16$. It is also in this regime ($\beta \geq 16$) where we find
the higher order (three or more) intercommutations. Further work is
needed to understand if the window closes for $\beta > 64$. We see
fewer multiple intercommutations, but this could be simply due to the
limitations of the dynamical range.
\\

As $\beta$ increases, we find a lower critical velocity for double (or
multiple) reconnections, in agreement with \cite{Putter}, but for very
large $\beta$ we are unable to describe the angular dependence in
detail. The fit derived in \cite{Putter} works well for $\beta \leq 8$
and maybe even for $\beta = 16$ but not in the deep type-II
regime $\beta \geq 32$.
The interaction between the magnetic cores produces a torque that
tends to anti-align the string segments, so the actual collision angle
is higher than in the initial configuration, but this is only true
around the collision point, and this portion of string quickly
disappears. In fact, due to this antialignment, the collision results
in the annihilation of a larger segment of string so in fact the
strings behave as if they that had collided with a {\it smaller}
angle. The simplest way to model the bending of the string, as a
sinusoidal excitation, is shown in fig. \ref{string_bending2D} and
gives the angle $\xi$ in terms of $\alpha$: $\xi/2 = \alpha -
\pi/2$. Using $\xi/2$ instead of $\alpha/2$ in eq.  \ref{w_crit}, we
find a one-parameter fit to our data with initial collision angles
higher than 120 degrees.  The fit is shown with a dashed line in
figure \ref{vcrit_panels}. \\

Even for lower collision angles, the fit derived in \cite{Putter},
with or without an adjustment to include the torque between the
colliding strings, does not fully describe the angle dependence of the
critical velocity. Our data (see fig. \ref{vcrit_panels}) is possibly
better described in the deep type-II regime by a curve of opposite
concavity with a plateau between collision angles of $80^0$ and
$120^0$. A more detailed study is necessary to
understand the angle dependence of the critical velocity.\\


We also simulated two dimensional vortex-antivortex head-on collisions
in an attempt to understand the new 3D effects.  We confirmed the
result of Myers et al. \cite{MyersRebbiStrilka} that for high enough
collision speed the vortex-antivortex pair reemerges some time after
the collision.  The critical speed for reemergence goes down with
increasing $\beta$, and for $\beta = 32$ is already lower than $v \sim
0.7$. On the other hand, for $\beta < 8$ the velocity needed for the
vortex-antivortex pair to reemerge is so high ($v \geq 0.98$) that we
only see a (bound) radiating state. In particular, for $\beta \leq 4$
Myers et al. report the backscatter of the vortex-antivortex pair
whereas we always see a radiating bound state.  This is however not
inconsistent with \cite{MyersRebbiStrilka} given
their much smaller dynamical range (see fig. \ref{2D_panels}). So we see a transition between reemergence in forward direction vs a bound radiating state, we never see backscatter.\\

Further work is needed to determine the critical $\beta$,
in 2D head-on v-av collisions, that distinguishes the regime where a
bound radiating state forms from the regime where the v-av pair
reemerge as if they had passed through.  We can locate this transition
somewhere around $6.2 \leq \beta \leq 6.4$ but the main problem in
determing this critical $\beta$ is the high collision speed ($v >
0.98$) needed for reemergence, which leads to very bad resolution.\\

Our expectation is that the 2D radiating bound states
at low $\beta$ should roughly correspond to the 3D blob, and that 2D
forward reemergence should correspond to the 3D loop. For not too
large $\beta$ (in particular below 6.2) the critical velocity for
passing through in 2D is so high ($ > 0.98c$ ) that we do not probe it
in either the 2D or 3D simulations, we always see a radiation blob. As
$\beta$ increases, the critical velocity for passing through in 2D
goes down and at high $\beta$ (in particular $\geq $32) the critical
velocity for forward reemergence is so low ($< 0.7 c$) that all the 3D
simulations are in this regime, and all show  loop formation.   \\

Although suggestive, our interpretation of the results is not
conclusive. To confirm this picture one should study the transition
region $8 \leq \beta \leq 32$, $0.7 \leq v \leq 0.95$ both in 2D and
3D and verify the extent of this correlation (it is important to note
that we have no data between $\beta = 8$ and $\beta = 16$). Also, we
do not necessarily expect a precise, one-to-one correspondence in the
critical values of $\beta$ and $v$ because the energy requirements to
reform a vortex--antivortex pair in 2D are different from those needed to form a loop  in 3D.\\









The effect on the cosmological signatures of strings is hard
to predict, as the stronger core interactions in the deep type-II
regime affect energy loss mechanisms in several, competing,
ways\cite{us}.
In general, we expect a relative enhancement of the radiation
contribution from kinks and reconnections at the expense of cusps
(suppressed by the kinks \cite{KibbleTurok,Garfinkle:1987yw}) and loops
(suppressed by the lower critical
velocity for double reconnection). \\

Regarding kinks, a new feature is the presence of kink trains
resulting from multiple reconnections.  These are rare, but once a
kink train is formed its decay time is comparable to that of a single
kink, and because of its microscopic size (it is only a few core
widths in length) it is very unlikely to be disrupted by
intercommutation with another string segment. We conclude that the
small scale structure of strongly type-II Abelian Higgs string
networks could be somewhat more clustered than the predictions based
on the Nambu-Goto approximation with $P=1$
\cite{KibbleCopeland,PolchinskiRocha,CopelandKibble_kinks}, although nowhere near the proliferation of kinks expected in a network with junctions\cite{Binetruy:2010bq}. \\

Regarding reconnections, they are enhanced in two ways. First, while
we confirm that strongly type-II ANO strings {\it always reconnect}
($P=1$), an {\it effective} intercommutation probability $P_{eff} \leq 1$
due to multiple reconnections will still lead to denser networks and
therefore more reconnections. Second, most of these will be at low
velocity, and we have argued that antialignment plays a role even in
relativistic speed collisions, so we expect a strong effect in low
velocity collisions.  So we would expect stronger and more frequent
bursts of radiation and cosmic rays than for other string types (lower
$\beta$ and also superstrings) where reconnection bursts are always
negligible or subdominant \cite{DamourVilenkin01,JacksonSiemens}.\\

Further work is needed to understand these effects quantitatively, and
how they affect the cosmological bounds. But the upshot of the work
presented here is that core interactions are expected to cause
significant differences with respect to the predictions from both
Nambu--Goto strings and field theory Abelian Higgs strings in
the Bogomolnyi limit.\\

Finally, our results confirm once again that Abelian Higgs strings
{\it always reconnect}, even at ultrarelativistic speeds ($P=1$);
unlike for other types of defects
\cite{MyersRebbiStrilka,Giblin:2010bd}, and against naive expectations,
the only way in which strings can pass through each other appears to
be by an even number of reconnections. \\



\begin{acknowledgments} We are grateful to Roland de Putter, Mark
  Jackson, Paul Shellard and Jon Urrestilla for discussions. Work
  supported by the Netherlands Organization for Scientific Research
  (NWO) under the VICI programme. We acknowledge partial support by
  the Consolider-ingenio programme CDS2007-00042.
\end{acknowledgments}

\bibliography{bib}

\end{document}